\documentclass[
amssymb,aps,
prd,nofootinbib,floatfix,11pt,
]{revtex4}%
\pdfoutput=1

\usepackage{amsmath,amssymb,epsf}
\usepackage{graphicx,color}

\usepackage{hyperref}

\usepackage{amssymb}
\usepackage{amsmath}
\usepackage{mathrsfs}
\usepackage{graphics}
\usepackage{epsfig}
\usepackage{graphicx}                   
\usepackage{subfigure}
\usepackage{bm}                              
\usepackage{nicefrac}
\usepackage{soul}

\usepackage{float}
\usepackage{placeins}
\usepackage{caption}
\usepackage{dcolumn}              

\usepackage[T1]{fontenc} 

\definecolor{cblue}{RGB}{100,5,255}
\definecolor{cred}{RGB}{255,50,40} 
\definecolor{cgreen}{RGB}{1,100,0} 

\def\lsim{\mathrel{\rlap{\lower4pt\hbox{$\sim$}}
    \raise1pt\hbox{$<$}}}                
\def\gsim{\mathrel{\rlap{\lower4pt\hbox{$\sim$}}
    \raise1pt\hbox{$>$}}}            

\newcommand\beq{\begin{eqnarray}}
\newcommand\eeq{\end{eqnarray}}

\begin{document}

\renewcommand{\theequation}{\arabic{section}.\arabic{equation}}
\renewcommand{\thefigure}{\arabic{section}.\arabic{figure}}
\renewcommand{\thetable}{\arabic{section}.\arabic{table}}


\title{\Large \baselineskip=10pt  A paradigm of warm quintessential inflation and production of relic gravity waves}

\vskip 2cm

\author{$^1$M.~R.~Gangopadhyay, $^{2,3,4}$Shynaray~Myrzakul, $^{3,5,6}$M.~Sami and $^7$M.~K.~Sharma}
\affiliation{
\it$^1$Centre For Theoretical Physics,Jamia Millia Islamia, New Delhi-110025, India.\\
\it$^2$ Ratbay Myrzakulov Eurasian International Centre for Theoretical Physics, Nur-Sultan, 010009, Kazakhstan.\\
 \it$^3$Center for Theoretical Physics, Eurasian National University, Nur-Sultan 010008, Kazakhstan.\\
 \it$^4$ Nazarbayev University,  Nur-Sultan, 010000, Kazakhstan\\
 \it$^5$International Center for Cosmology, Charusat University, Anand 388421, Gujarat, India\\
\it$^6$Institute for Advanced Physics and Mathematics, Zhejiang University of Technology, Hangzhou, 310032, China.\\
\it$^7$Department of Physics $\&$ Astrophysics, University of Delhi, Delhi-110007, India.}

\begin{abstract}\normalsize 
\normalsize \baselineskip=10pt 
We consider the  framework of quintessential inflation in the warm background which is caused by the dissipation of the scalar field energy density into relativistic degrees of freedom. The dissipation process has important consequences for the evolution both at the levels of background as well as perturbation and allows us to easily satisfy the observational constraints. Our numerical analysis confirms that the
model conforms to observational constraints even if the field-radiation coupling  coupling strength is weak. In the warm background, we investigate the post inflationary evolution of relic gravity waves produced during inflation whose amplitude is constrained by the Big Bang Nucleosynthesis constraint. Further, we investigate the effect of  coupling on the amplitude of gravity waves and obtain the allowed phase space between the model parameters. The mechanism of quintessential inflation  gives rise to blue spectrum of gravitational wave background at high frequencies. We discuss the perspectives of detection of the signal of relic gravity waves by future proposed missions. Improvements of their sensitivities in the high frequency regime, in future, might allow us to probe the blue tilt of the spectrum caused by the presence of kinetic regime in the underlying framework of  inflation.
\end{abstract}

\maketitle

\vspace{0.001in}


\baselineskip=15.4pt

\vspace{1cm}

\maketitle



\section{Introduction}\label{sec:intro}
\setcounter{equation}{0}
\setcounter{figure}{0}
\setcounter{table}{0}
\setcounter{footnote}{1}
The standard paradigm of cosmic inflation has been proven consistently over the years, by large-scale observations like Planck and WMAP etc.~\cite{ade2016planck,aghanim2018planck,hinshaw2013nine}, as one of the most promising way to generate Gaussian as well as adiabatic perturbations with high accuracy, in addition to addressing the inherent inconsistencies of the standard model of Universe \cite{besset}. In the standard inflationary set up, a massive slow-rolling scalar field with a flat potential gives rise to (quasi)exponential cosmic expansion~\cite{linde,shinji-rev}. The field also assumes to be weakly coupled to other pre-existing matter fields such as radiation as the expansion causes them to redshift-away. Thus, the field in the standard inflationary framework is treated as a \textit{gauge singlet} and also assumes that the temperature remains negligible all the way till the reheating commences, hence known as ``Cold Inflation'' (CI).

Inspite of its overwhelming success, the standard inflationary paradigm has been challenged by the recently proposed swampland conjectures (SC) which are based on the idea that the stable de-Sitter (dS) vacuum is quite difficult to construct in the quantum theory of gravity \cite{obied,kallosh,agrawal}. Also, the conditions imposed by them are formidable to satisfy and hence one requires to consider other inflationary frameworks. It has been shown that, to satisfy such a condition, one either has to invoke the braneworld scenario~\cite{mrg1,brahma1} or more interestingly, these conditions can be fulfilled in a rather realistic dynamical realisation known as ``Warm Inflation'' (WI)\cite{suratna,berera,dimop-owen,shojai}.

The basic concept of WI scenario is based on the fact that the scalar field, instead of acting as a gauge singlet, dissipates into radiation via its coupling with it \cite{berera,branden,hall-moss}. The dissipation takes place throughout the inflation and causes it to end when the slow roll is violated. Also, due to the presence of radiation, the temperature during inflation is generally non-zero and could even be comparable to or greater than the Hubble parameter \cite{oliviera,graham}. If the temperature is lower (or greater) than the Hubble parameter, quantum (or thermal) fluctuations dominates the universe. Furthermore, as inflation proceeds, the radiation energy density also increase giving rise to the ``graceful-exit'' \cite{cerezo,herrera-vid,berera2}.
It has also been shown that the WI model not only resolves theoretical issues some of them as mentioned above, but it also resurrects the inflationary potentials otherwise ruled out in the CI paradigm by the observations. \cite{panotopoulos,bartrum,bastero-gil,arya2}.

In this paper, we investigate a quintessential inflationary model in the thermal background. The model includes a generalized exponential field potential which remains flat during inflation and becomes steep in the post inflationary era \cite{hossain2015unification,geng2017observational,hossain}.
In this case, the field enters the kinetic regime soon after inflation ends and continues to be there even after it is taken over by the background
and freezing is reached due to Hubble damping.~Thereafter, as the background energy density becomes comparable to field energy density, field evolution commences again and ultimately catches up with the background and tracks it till the additional late time features in the field potential become operative.
In this case, one needs to invoke alternative reheating mechanism which have their limitations. The paradigm of warm inflation, considered here, not only serves as a source of reheating but also adds generic features to inflationary framework itself. In this set up, during inflation, the field and radiation both interact with each other via the temperature-independent coupling parameter and assumes both of them to evolve independently after the commencement of the kinetic regime. The model very well describes the thermal history of the universe and also unifies both early and late-time phases of acceleration. It has been shown in \cite{geng2017observational} that the model satisfies the Cosmic Microwave Background (CMB) observations only if the field potential is very steep at the end of inflation. 

In this paper, our main goal is to consider inflation in a warm background and to check whether  the coupling can turn the field potential less steeper,  compared to the standard inflationary framework, subject
to various observational requirements. To this effect, we examine the model under  observational constraints on: (\textit{i}) Scalar perturbation spectra, (\textit{ii}) baryon-to-entropy ratio and (\textit{iii}) relic gravitational waves (GW). 

The standard  preheating mechanism is not applicable to the framework of quintessential inflation. One of the alternatives is provided by the gravitational particle production which is a universal process. However, the latter is inefficient and might give rise to violation of nucleosynthesis constraint due to relic gravity waves. Instant preheating and curvaton mechanism have their limitations\footnote{As for instant preheating, in absence of shift symmetry, one needs to invoke adhoc assumption to throw out extra term in the Lagrangian containing two extra fields whereas curvaton is implemented using an extra field.}. The framework of warm inflation does not require the presence of extra fields, it rather assumes coupling between the inflaton and radiation. In the discussion, to follow, we briefly, mention the underlying mechanism to be applied to quintessential inflation in the subsequent sections.

\section{Model of warm inflation and inflationary parameters}
\setcounter{equation}{0}
\setcounter{figure}{0}
\setcounter{table}{0}
\setcounter{footnote}{1}
\label{sec:potrec}
In the framework of warm inflation, the dissipation of scalar field into light degrees of freedom occurs rapidly than the Hubble expansion rate; the field-radiation interaction parameter $Q= \Upsilon /3H $ (where $\Upsilon$ is a dissipation coefficient), in turn, gives rise to damping of the field evolution and also flatten its potential,

\begin{eqnarray} {\label{slow-rolleqns}}
3H(1+Q)\dot{\phi} &\simeq& -V_{,\phi},  \quad \mbox{where} \quad \cdot \equiv \frac{d} {dt} \\
\rho_r &\simeq& \frac{3}{4}Q \dot\phi^2 \ ,
\end{eqnarray}
where $H$ is the Hubble parameter, $V$ is the field potential and $\rho_r$ is the energy density of the radiation. Here, we assume that field and radiation remains quasi-constant, which implies that, $\ddot{\phi}\ll \dot{\phi}$, $\ddot{\phi} \ll V_{,\phi}$ and $\dot{\rho_r} \ll \rho_r$. The inflationary regime is characterized by two slow-roll parameters $\epsilon$ and $\eta$, 
\begin{equation} \label{slow-roll param}
\epsilon = \frac{1}{2}M_{pl}^2  \left(\frac{V_{,\phi}}{V}\right)^2 \quad \mbox{and} \quad \eta = M_{pl}^2 \frac{V_{,\phi \phi}}{V} \ .    
\end{equation}
Due to the presence of coupling $Q$, the slow-roll condition now requires, $\epsilon,|\eta| \ll 1+Q$. Note that, one can now have, $\epsilon,|\eta| \geq 1$ which is generally prohibited in standard CI. The presence of coupling enhances damping
similar to an effect in RS brane worlds in high energy regime \cite{mrgbw, sukbw}.
%
%

In the framework quintessential inflationary,  after the inflation ends, the field enters  the kinetic regime and stays there for quite long before the commencement of radiative regime. The  field might further sources the late time cosmic acceleration thanks to its non-minimal coupling with massive neutrino matter \cite{geng2017observational}. In this scenario, a field potential which can source both the early and late time cosmic acceleration is given by,
\begin{equation} \label{Vphi}
    V(\phi) = V_0 \, \exp\left[-\lambda\left(\frac{\phi}{M_{pl}}\right)^n\right] \  .
\end{equation}
For this choice of potential, the time-derivative of field can be written as
\begin{equation} \label{phit}
\dot{\phi}(t) \simeq \frac{\lambda \,  n \, V_0 \, \phi ^{n-1} \exp({-\lambda  \phi ^n})}{3 H(1+ Q)} \ .
\end{equation}
Using Eq. (\ref{slow-roll param}), the slow-roll parameters can be worked out as
\begin{eqnarray} \label{eps,eta}
    \epsilon &=& \frac{1}{2}n^2 \left(\frac{\phi}{M_{pl}}\right)^{2n-2}\lambda^2 \ , \nonumber \\
    \eta &=& n \left(\frac{\phi}{M_{pl}}\right)^{n-2} \lambda\left(1+n\left(\lambda \left(\frac{\phi}{M_{pl}}\right)^n -1\right)\right) \ .
\end{eqnarray}
Inflation ends when either $\epsilon(\phi=\phi_{end})=1+Q$ or $\eta(\phi=\phi_{end})=1+Q$, which gives the expression of field,
\begin{equation} \label{phie}
    \phi_{end} = \left(\frac{2(1+Q)}{n^2\lambda^2}\right)^\frac{1}{2n-2} M_{pl} \ .
\end{equation}
Since the radiation density, at this epoch,  is negligible compared to the field energy density ($\rho_r\ll V(\phi)$), one can approximate $H$ as
\begin{equation} \label{H}
H^2 \simeq \frac{V(\phi)}{3 M_{pl}^2} = \frac{V_0 \, \exp\left[-\lambda\left(\frac{\phi}{M_{pl}}\right)^n\right]}{3 M_{pl}^2}\ '
\end{equation}
see Refs\cite{Lima:2019yyv,Das:2020xmh,Das:2020lut} on the related theme.
Unlike cold inflation, the warm inflation is characterised by its warmness or temperature $T$. The quasi-constant radiation density instantly produces a close to a thermal-equilibrium configuration in space, thus gives rise to nearly same temperature everywhere. And at temperature $T$, the energy density of relativistic particles is given by,
\begin{equation} {\label{rhor-T}}
\rho_r \simeq g_{\ast}\frac{\pi^2}{30}T^4 \ ,
\end{equation}
where $g_\ast$ is the relativistic degree of freedom. In this paper, we shall focus on the analysis of inflationary observables of the two successful realization of the dissipation forms known as the {\bf Linear ($\Upsilon = C_T T$)} and {\bf Cubic ($\Upsilon = C_\phi \frac{T^3}{\phi^2}$)} in the WI paradigm\cite{Bastero-Gil:2016qru,Bastero-Gil:2019gao}. By using the relation $\Upsilon=3HQ$, one finds that
\begin{equation} \label{T_lin_cubic}
    T =
    \begin{cases}
    \frac{3HQ}{C_T}, \quad \text{ {\bf Linear case}} \  \\
    \left(\frac{3HQ \phi^2}{C_\phi}\right)^{2/3}, \quad \text{ {\bf Cubic case}} \ . \\
    \end{cases}
\end{equation}
Then for the linear case, we can express $Q$ in terms of $\phi$ using Eqs. (\ref{slow-rolleqns}), (\ref{H}), (\ref{rhor-T}) and (\ref{T_lin_cubic}), as
\begin{equation} \label{Q-phi_lin}
Q=\frac{6 \exp(-\lambda \,\phi^n) \,g_\ast\, \phi^{2-2n}\, \pi^2\, Q^4\,(1+Q)^2 \,V_0}{5 \,C_T^4\, n^2 \, \lambda^2} \ .
\end{equation}
By differentiating $Q$, with respect to $N := \ln a$, we find
\begin{equation} \label{dqdn_lin}
\frac{dQ}{dN}= -\frac{6 \, g_\ast\, n \,\pi^2\, V_0\, \lambda\, \phi^n\,(2n-2+n\lambda \phi^n)\,Q^4\,(1+Q)}{5\, C_T^4 \,\exp(\lambda \,\phi^n)\,n^2\,\lambda^2\, \phi^{2n}-12 \,g_\ast \,\pi^2 \,V_0 \,\phi^2\, Q^3\,(1+Q)\,(2+3Q)} \ ,
\end{equation}
where we have also used Eq.(\ref{phit}). Similarly, for the cubic case,
\begin{equation} \label{Q_cubic}
Q=\frac{2 \exp(\lambda\,\phi^n /3)\, g_\ast\, \phi^{\frac{14}{3}-2n}\, \pi^2 \,Q^{4/3} \,(1+Q)^2}{5\,\times 3^{1/3}\,C_\phi^{4/3}\, n^2 \,V_0^{1/3}\,\lambda^2} \ ,
\end{equation}
and hence
\begin{equation} \label{dqdn_cubic}
\frac{dQ}{dN}= -\frac{8\times3^{2/3}\exp(\lambda\,\phi^n/3)\,g_\ast\,\pi^2\, Q^{4/3}\,(1+Q)\,\lambda\,\phi^{\frac{8}{3}+n}(14-6n+n\lambda\,\phi^n)}{4 \times 3^{2/3}\exp(\lambda \,\phi^n /3)\,g_\ast\, \pi^2\, Q^{1/3}\,(1+Q)\,(2+5Q)\,\phi^{14/3}-45\, C_\phi^{4/3}\,n^2\, V_0^{1/3}\,\lambda^2 \,\phi^{2n}} \ .
\end{equation}
From Eqs. (\ref{Q-phi_lin}) and (\ref{Q_cubic}), one can express field $\phi$ in terms of model parameters. Then by equating $\phi$ with Eq.(\ref{phie}), one can eliminate $Q_{end}$ in favour  of other parameters. As for the number of e-folds $N$, we invert the Eq. (\ref{dqdn_lin}) and integrate $Q$ from its pivot value i.e. $Q_\ast$ to $Q_{end}$. Equating $N$ from its expected value $\in[50,70]$ one can now also express $Q_\ast$ in terms of model parameters. As a result, we find both $Q_\ast$ as well as $Q_{end}$ in terms of model parameters in the linear case. Similarly, the same procedure is carried out for the cubic case.

For the linear case: we depict the parametric dependence of $Q_\ast$ with $C_T$ for $N=50,60$ and $70$ \ref{Q-ct}, here we show that as $C_T$ increases, $Q_\ast$ also increases. Similarly, for the cubic case, we also find the same kind of evolution of $Q_\ast$ with $C_\phi$ (see fig. (\ref{cubic1})). In both cases, it shows that larger the dissipation of field into radiation, larger the field-radiation interaction and smaller the field evolution, or vice-versa. 


From Eqs. (\ref{slow-rolleqns}), (\ref{slow-roll param}) and (\ref{rhor-T}), the temperature, in general, can be expressed as
\begin{equation} \label{T}
T = \left(\frac{15}{g_\ast \pi^2} \frac{\epsilon \,Q\, V}{(1+Q)^2}\right)^{1/4} \ ,
\end{equation}
which at the end of inflation can be cast in the following form,
\begin{equation} \label{T}
T_{end} = \left(\frac{15}{g_\ast \pi^2} \frac{\,Q\, V_{end}}{1+Q}\right)^{1/4} \ ,  \quad \mbox{where} \quad V_{end} = V_0  \exp\left[-\lambda\left(\frac{2(1+Q)}{n^2\lambda^2}\right)^{\frac{n}{2n-2}}\right] \ ,
\end{equation}
where $V_{end}$ is the field potential at $\phi=\phi_{end}$.\\
In what follows, using the obtained estimates, we turn to the study of primordial perturbations in the framework under consideration.



\section{Primordial perturbations}
\setcounter{equation}{0}
\setcounter{figure}{0}
\setcounter{table}{0}
\setcounter{footnote}{1}
\label{sec:sect}
After inflation ends, any pre-existing quantum fluctuations in vacuum emerges as a scalar, vector and tensor perturbations (GW) in classical space-time background. Since, vector perturbations die out soon due to the space-time expansion, only scalar and tensor perturbations are of importance, their imprints  can be found in the CMB radiation as well as in the large-scale structure. 






\begin{figure} 
\centering{%
\includegraphics[height=3in,width=3.7in]{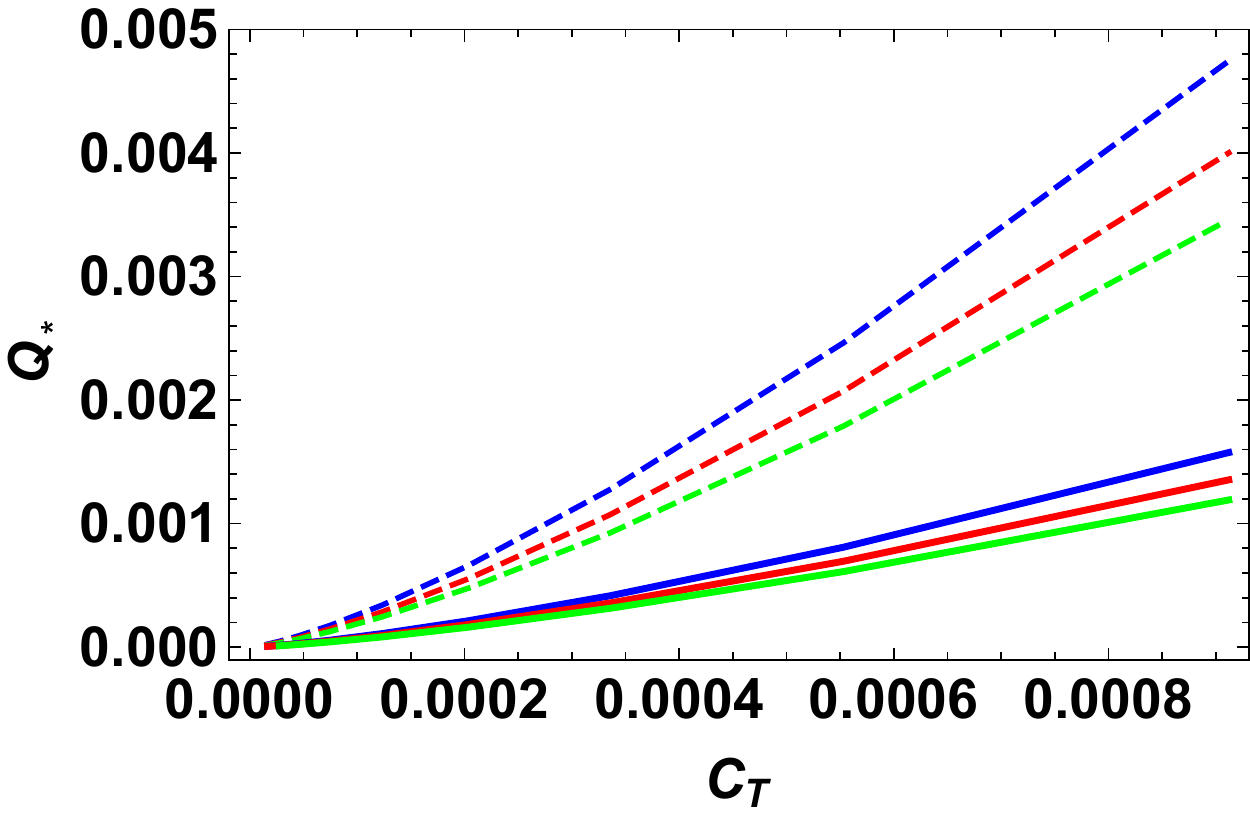}}%
\caption{\small Variation between $Q_\ast$ and $C_T \in [0,0.001]$ in case of linear dissipation regime. The blue, red and green color correspond to $N=50, 60$ and $70$, respectively, whereas solid and dashed lines represent the $n=4$ $\&$ $6$ cases, respectively.}
\label{Q-ct}
\end{figure}

\begin{figure} \label{ns-3d-lin}
\centering
\subfigure[]{
\includegraphics[height=2.2in,width=2.6in]{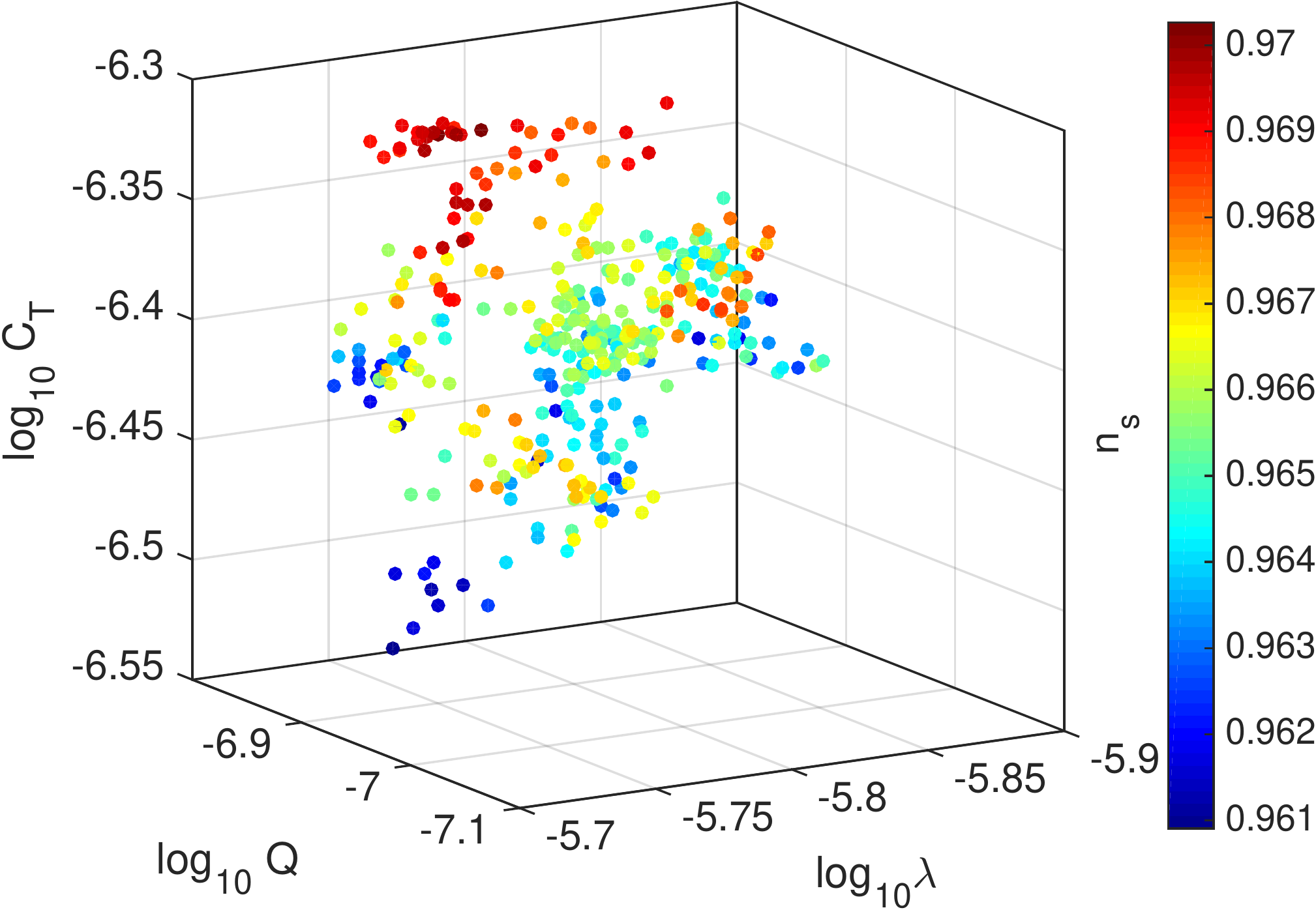}}%
\qquad \quad
\subfigure[]{%
\includegraphics[height=2.2in,width=2.6in]{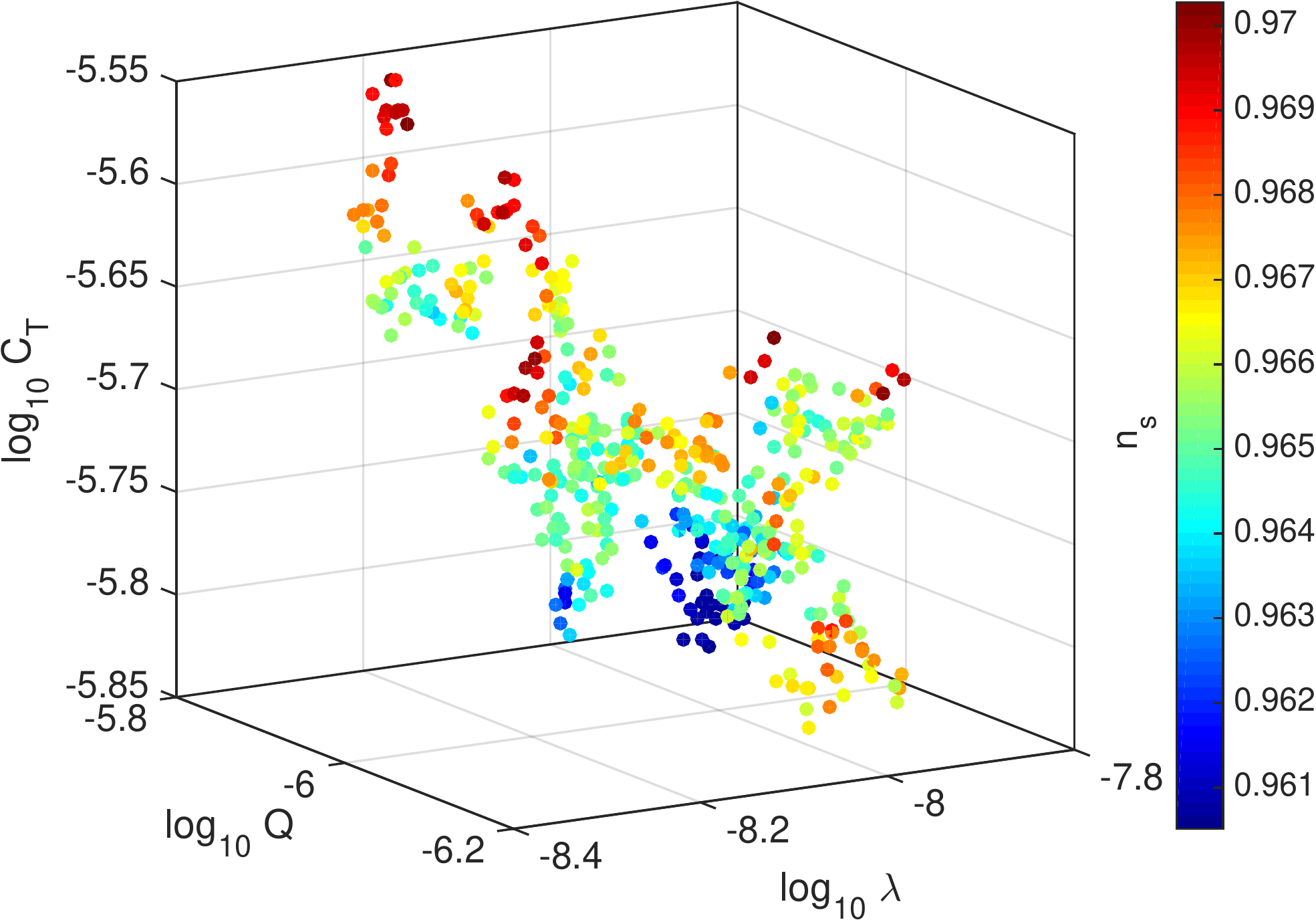}}
\caption{\small In case of linear dissipation regime, figure (a) and (b) shows a 3-dimensional allowed parametric region between $Q_\ast$, $\lambda$ and $C_T$ for the Planck2018+lensing+BK15+BAO \cite{aghanim2018planck} constraint on $n_s$ (shown in the bar) for cases $n=4$ and $6$, respectively.}
\label{linear3}
\end{figure} 

\begin{figure} 
\centering
\subfigure[]{%
\includegraphics[height=2.6in,width=2.6in]{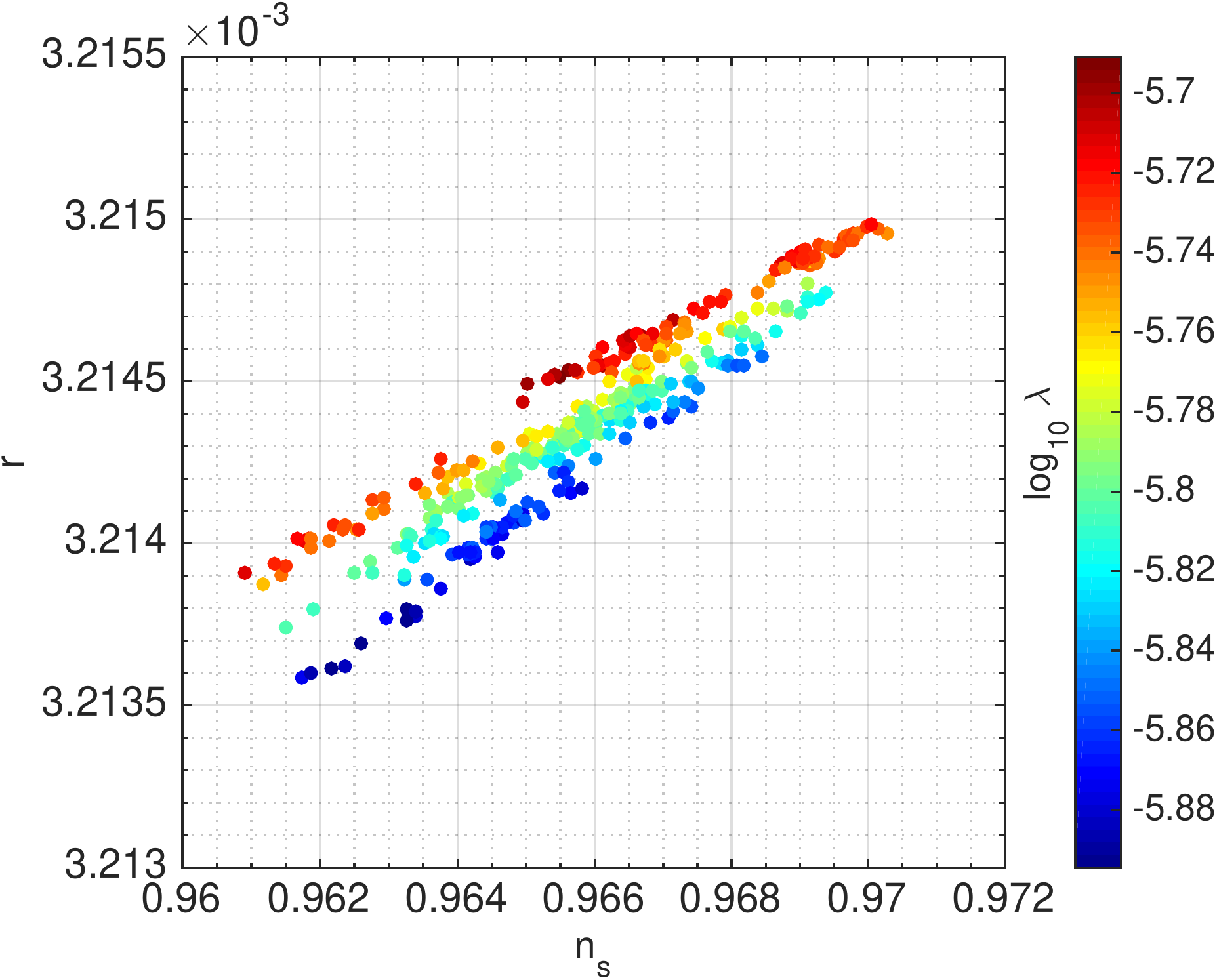}}%
\qquad \quad
\subfigure[]{%
\includegraphics[height=2.6in,width=2.6in]{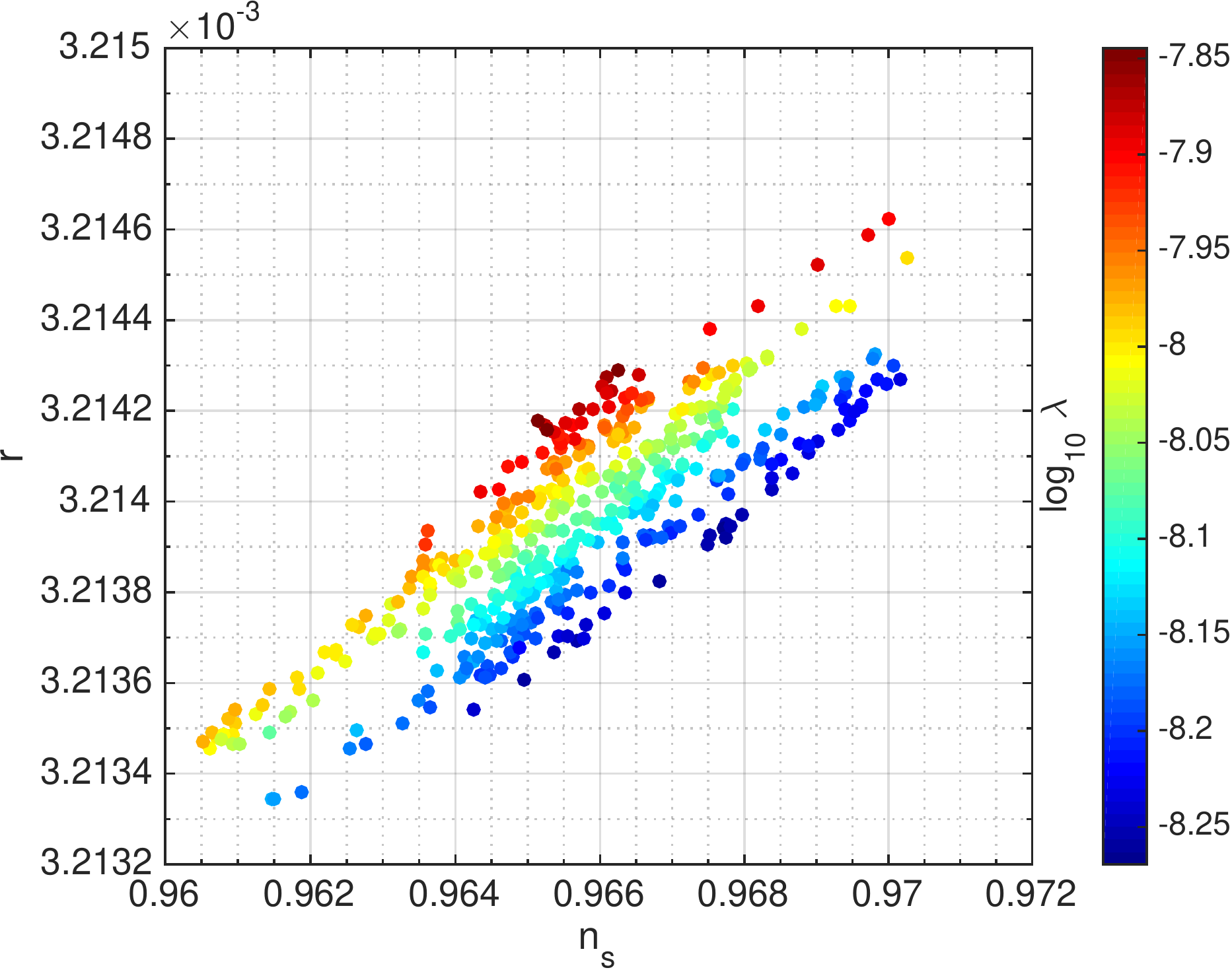}}%
\caption{\small In case of linear dissipation regime, figure (a) and (b) show a range of parameter $\lambda$ (shown in the bar) for cases $n=4$ and $6$, respectively, which is compatible with the Planck2018+lensing+BK15+BAO \cite{aghanim2018planck} constraint on $n_s$ and $r$.}
\label{nssr}
\end{figure}

\begin{figure}
\centering
\subfigure[]{%
\includegraphics[height=2.2in,width=2.6in]{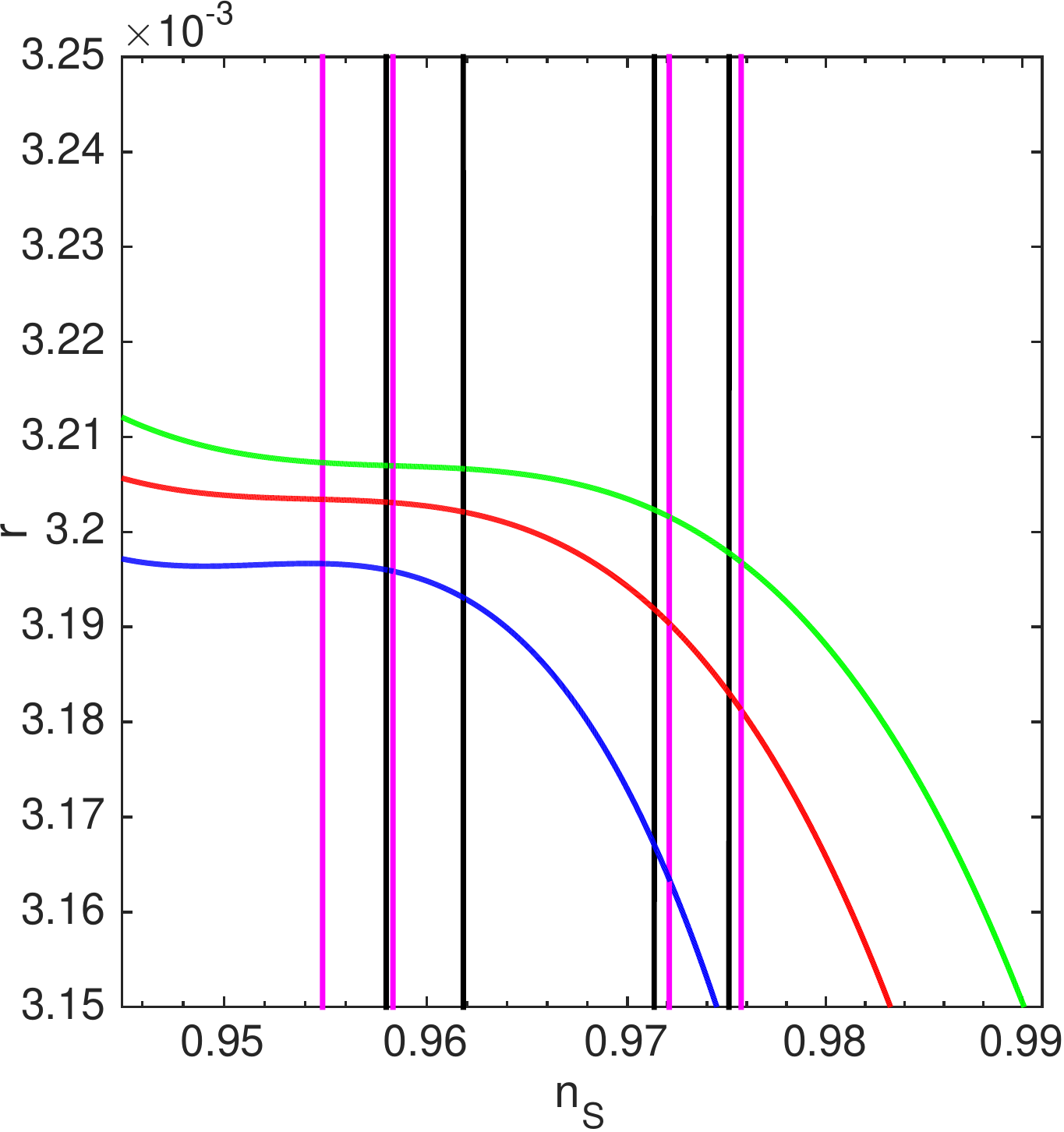}}%
\qquad \quad
\subfigure[]{%
\includegraphics[height=2.2in,width=2.6in]{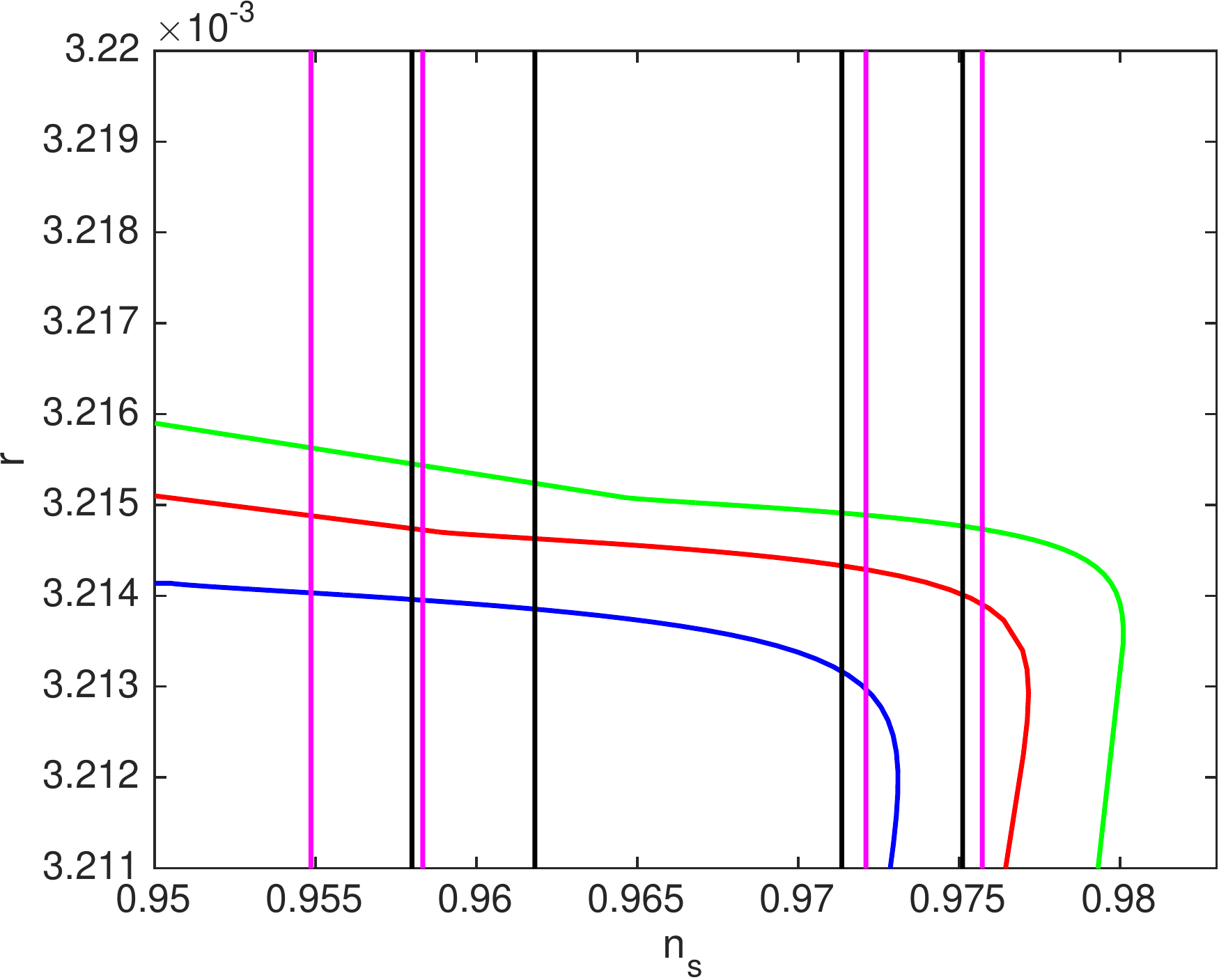}}
\caption{\small In case of linear dissipation regime, figure (a) and (b) represents plot between $n_s$ and $r$ for cases $n=4$ and $6$, respectively The magenta lines are for the Planck2018 TT, TE, EE+ low E+ lensing data, whereas the black solid lines correspond to Planck2018+lensing+BK15+BAO data. The pivot scale is fixed at usual $k=$ 0.002Mp$c^{-1}$. The green, red and blue curves depicts $N=50, 60$ and $70$, respectively.
}
 \label{ns-r-lin figure}
\end{figure}


\subsection{Scalar perturbations}
\begin{figure}[t]\label{Q-ct figure}
\centering
{%
\label{fig:first}%
\includegraphics[height=3in,width=4in]{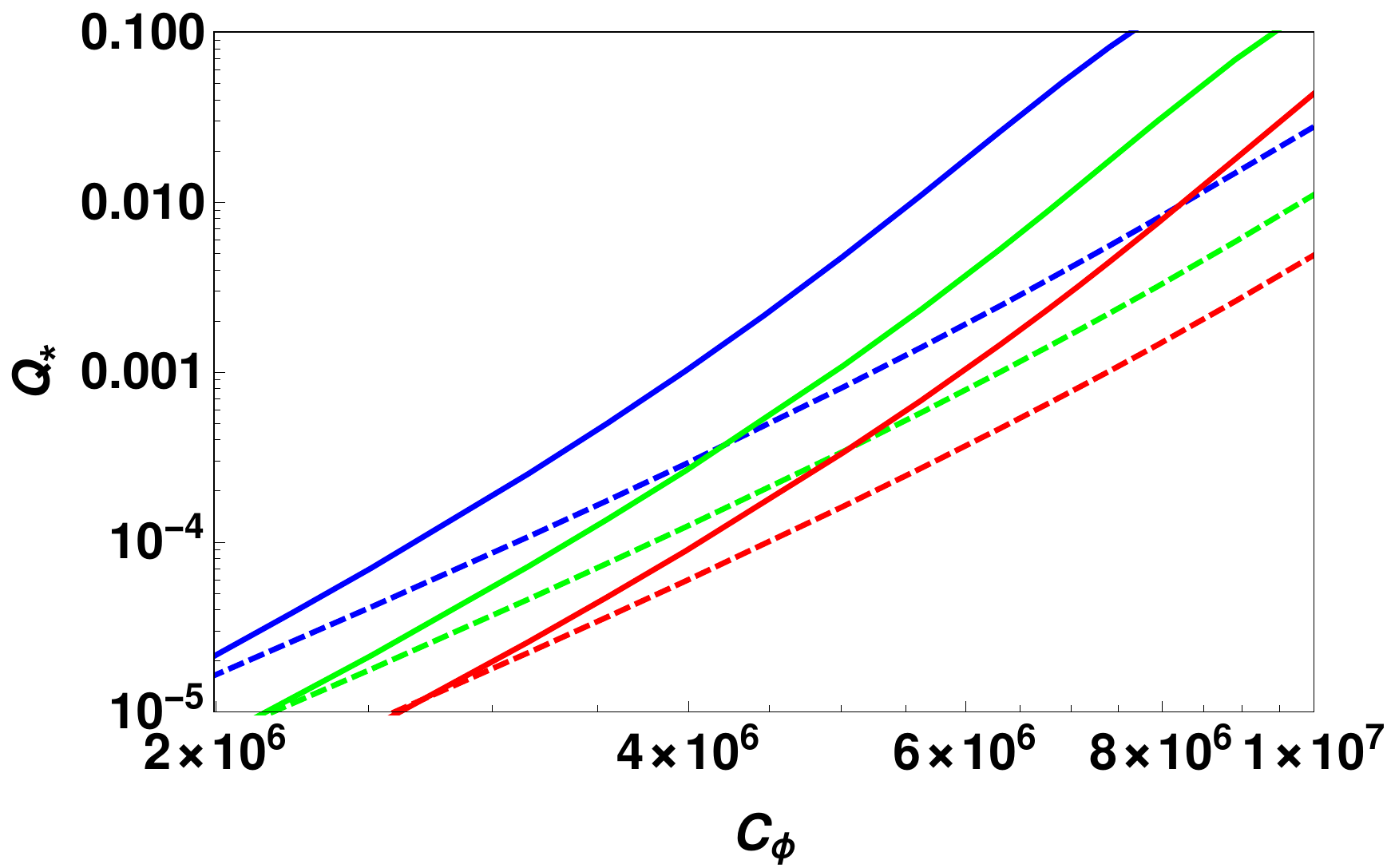}}%
\caption{Variation between $Q_\ast$ and $C_\phi \in [2\times 10^6,10^7]$ in case of cubic dissipation regime. The green, red and blue color represents $N=50, 60$ and $70$, respectively, whereas solid and dashed lines corresponds to the $n=4$ and $6$ case, respectively.}
\label{cubic1}
\end{figure}

\begin{figure}
\centering
\subfigure[]{%
\label{fig:first}%
\includegraphics[height=2.3in,width=2.7in]{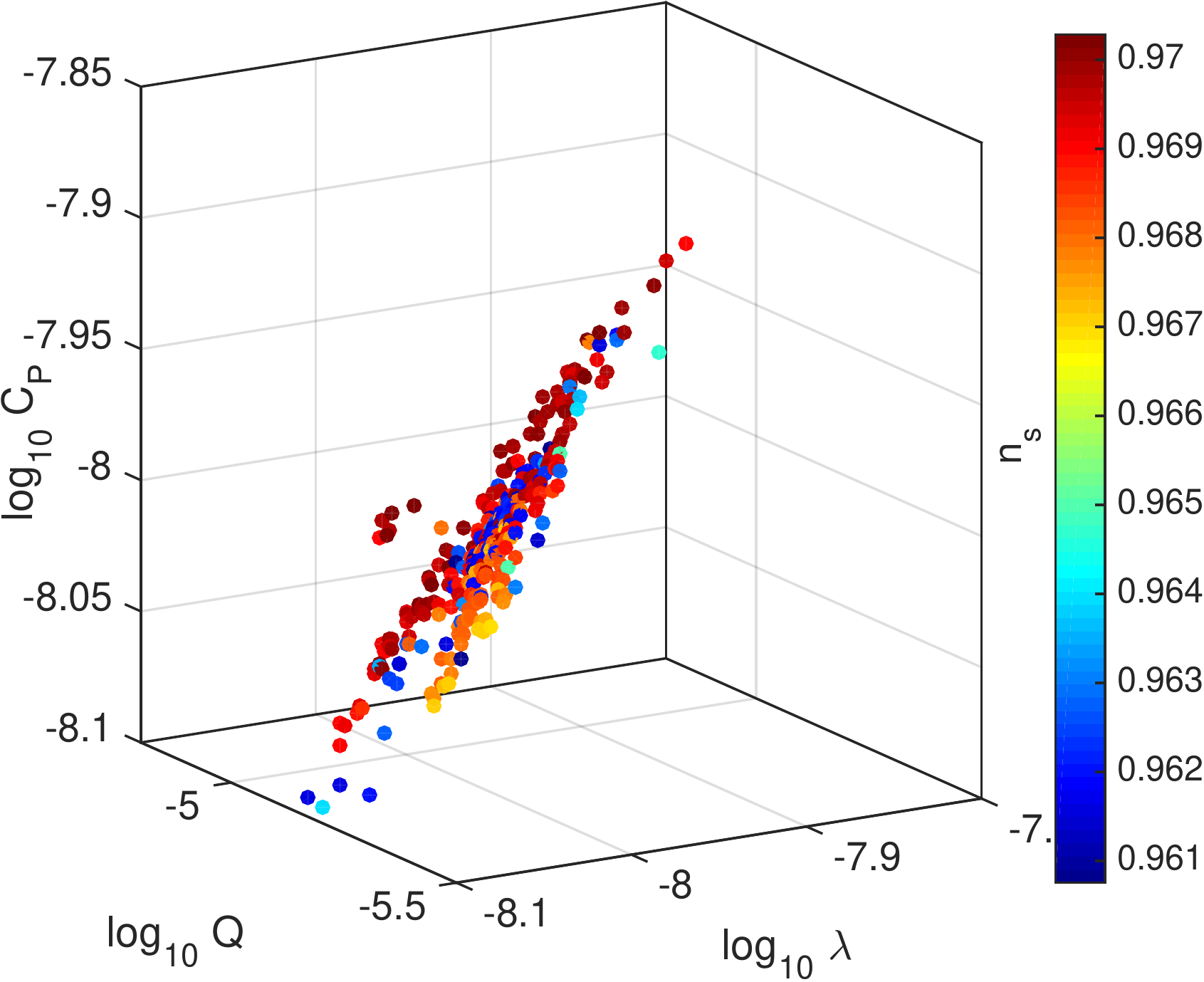}}%
\qquad \quad
\subfigure[]{%
\label{fig:second}%
\includegraphics[height=2.2in,width=2.6in]{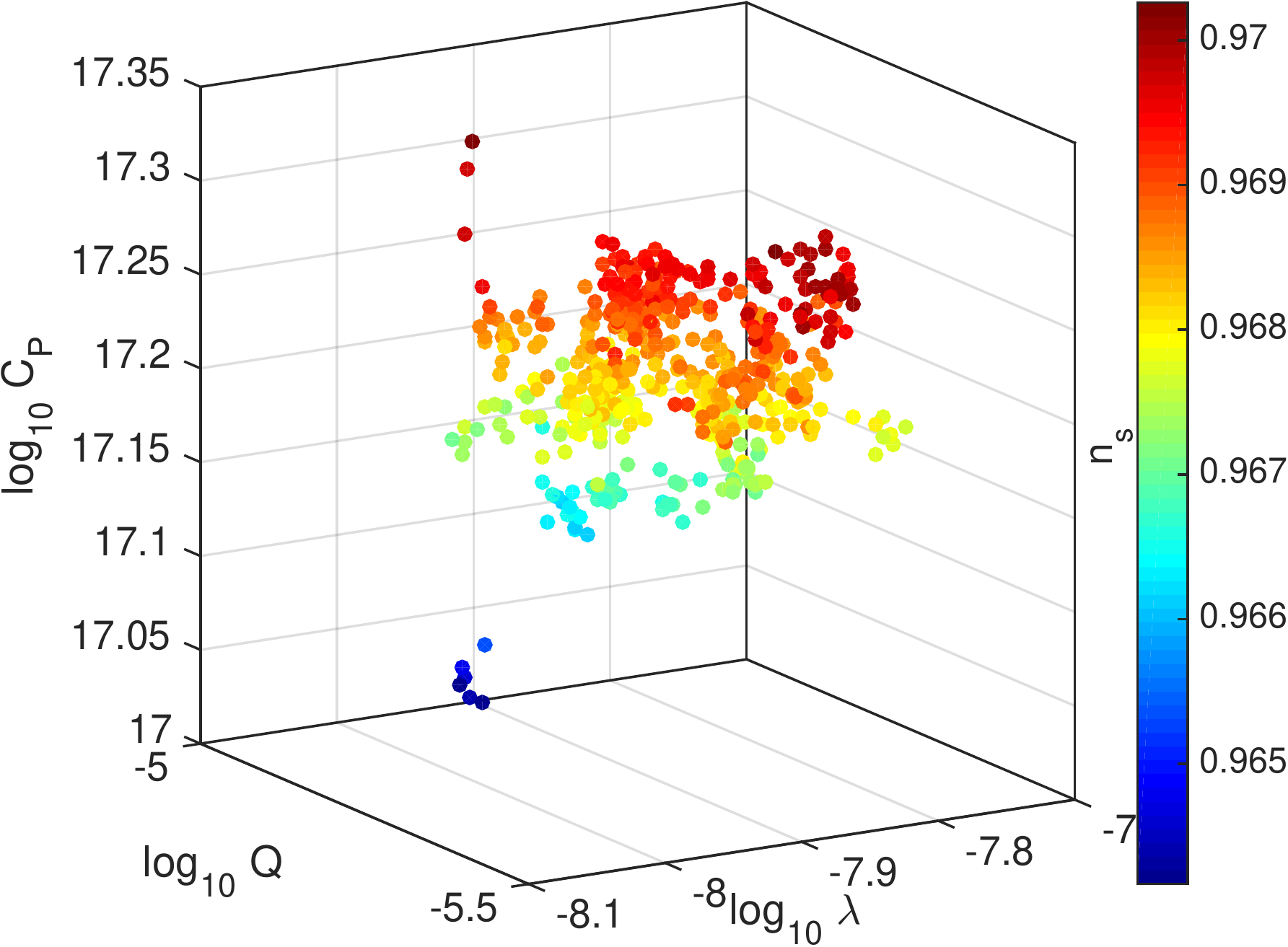}}
\caption{\small In case of cubic dissipation regime figure (a) and (b) shows a 3-dimensional allowed parametric region between $Q_\ast$, $\lambda$ and $C_\phi$ for cases $n=4$ and $6$, respectively. The region satisfies the Planck2018+lensing+BK15+BAO \cite{aghanim2018planck} constraint on $n_s$ (shown in the bar).}
\label{cubic3}
\end{figure}

\begin{figure} 
\centering
\subfigure[]{%
\label{fig:first}%
\includegraphics[height=2.2in,width=2.6in]{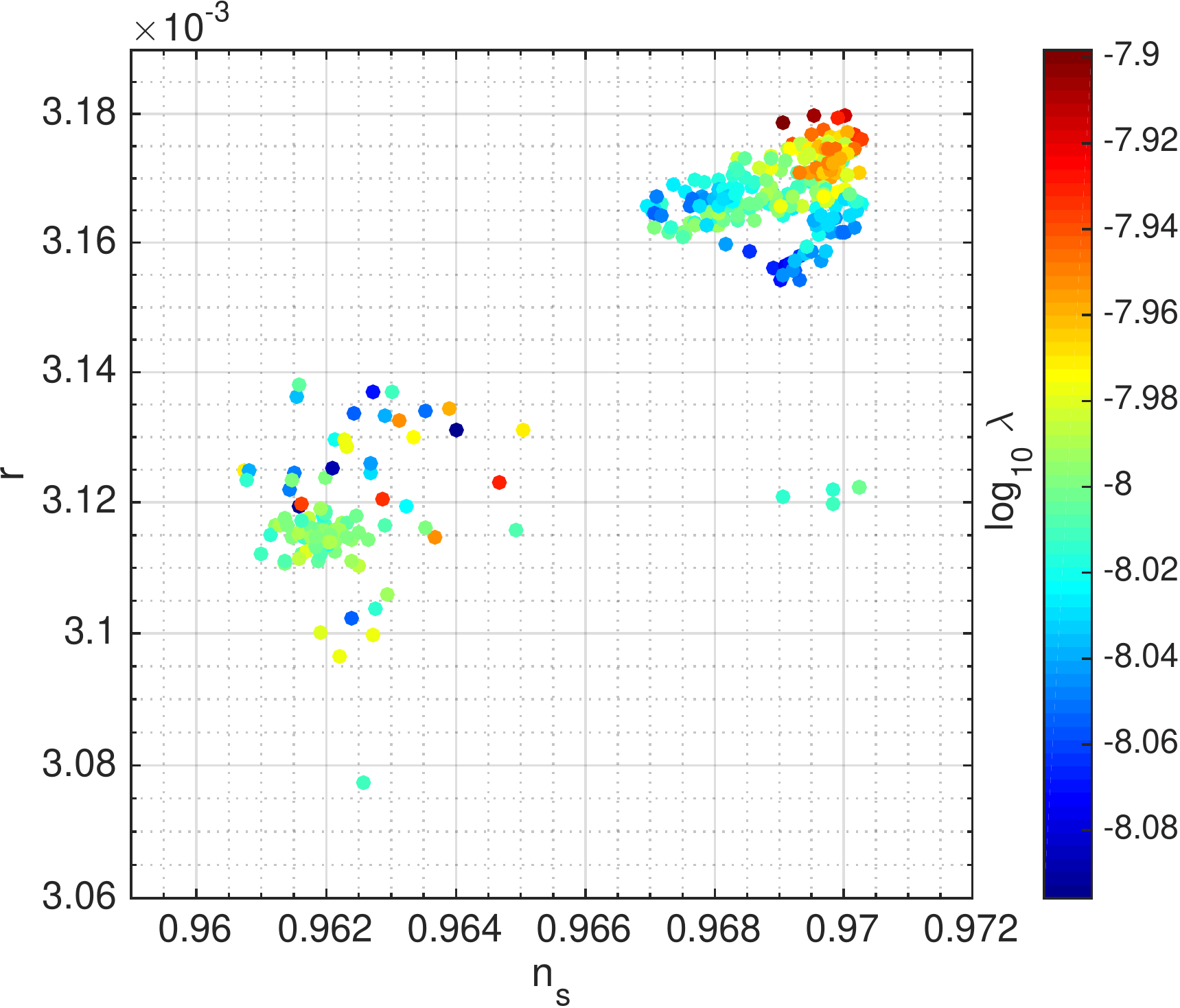}}%
\qquad \quad
\subfigure[]{%
\includegraphics[height=2.2in,width=2.4in]{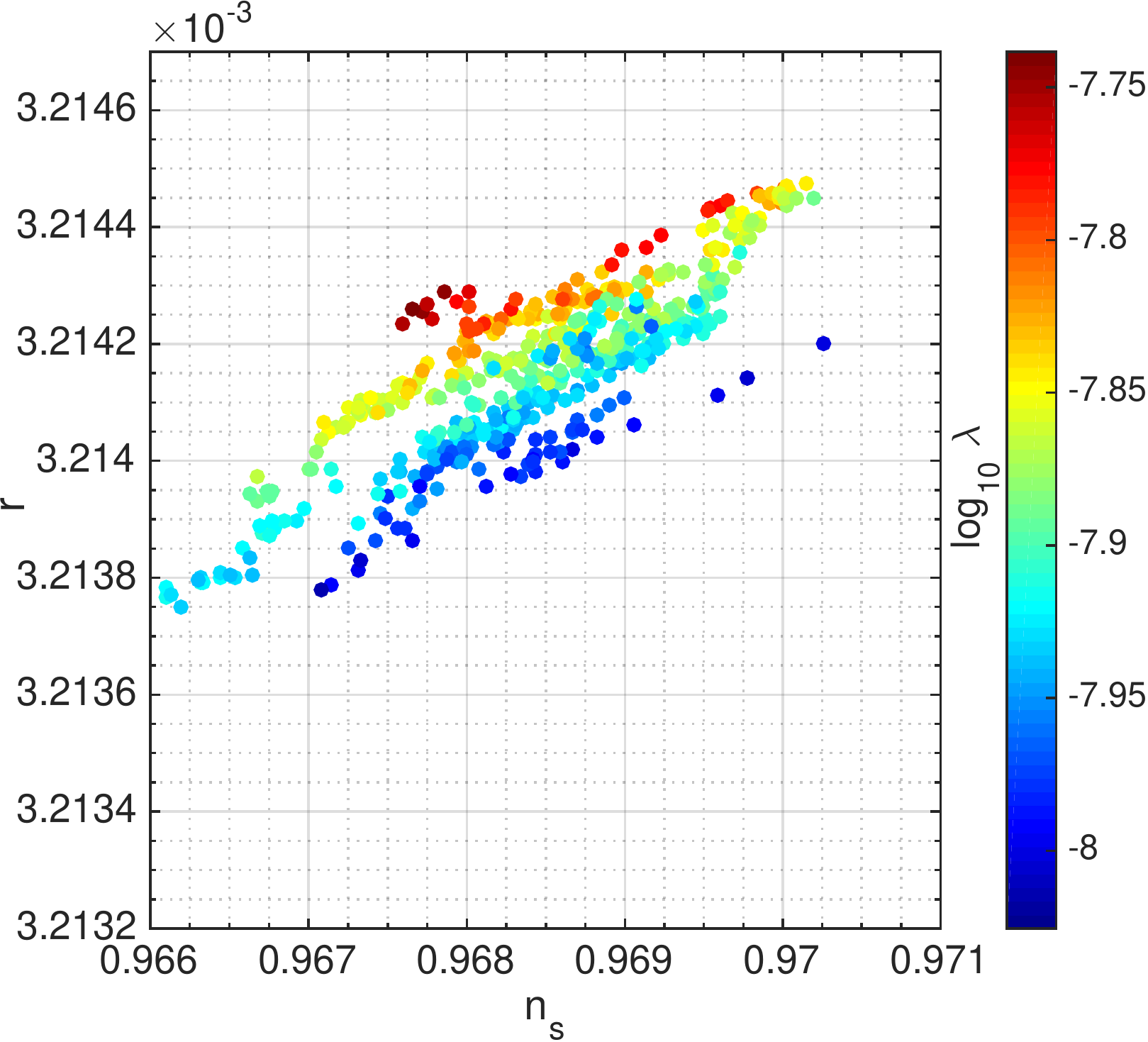}}
\caption{\small In case of cubic dissipation regime figure (a) and (b) shows a range of parameter $\lambda$ (shown in the bar) for cases $n=4$ and $6$, respectively, which is in compatible for the Planck2018+lensing+BK15+BAO \cite{aghanim2018planck} constraint on $n_s$ and $r$.}
\label{cubic2}
\end{figure}

\begin{figure} 
\centering
\subfigure[]{%
\includegraphics[height=2.2in,width=2.6in]{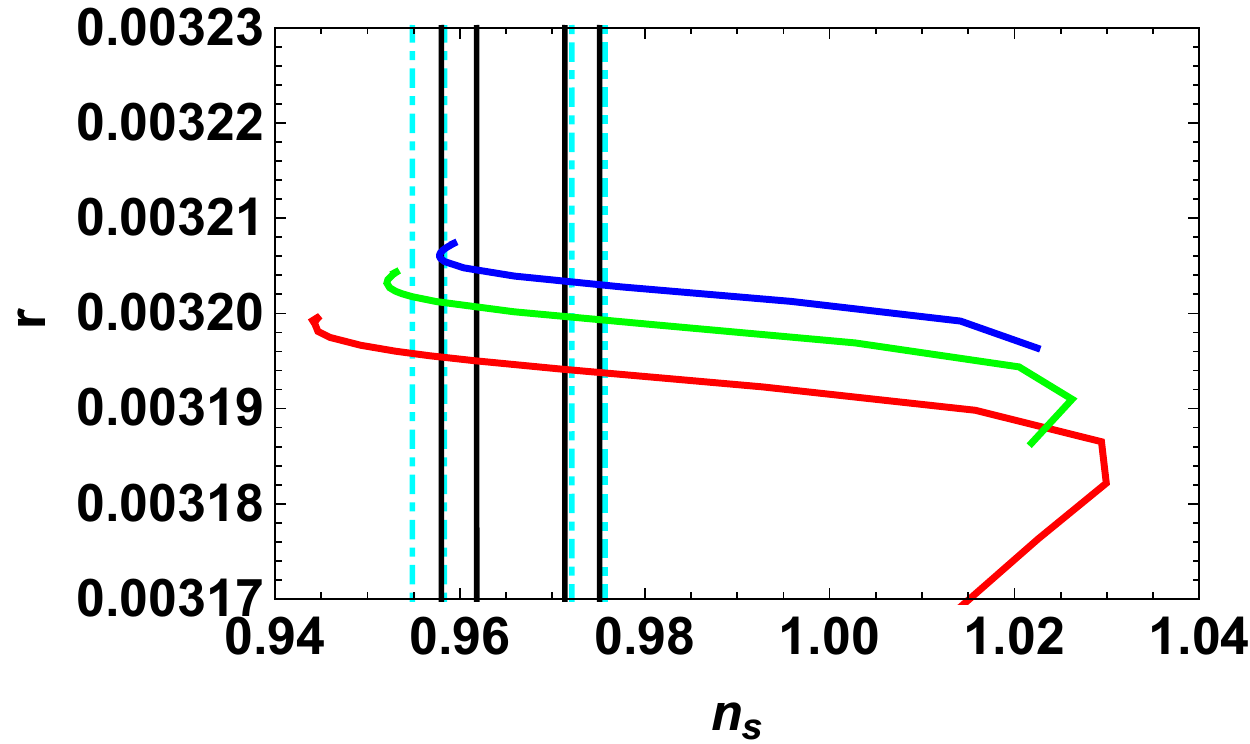}}%
\qquad \quad
\subfigure[]{%
\includegraphics[height=2.2in,width=2.6in]{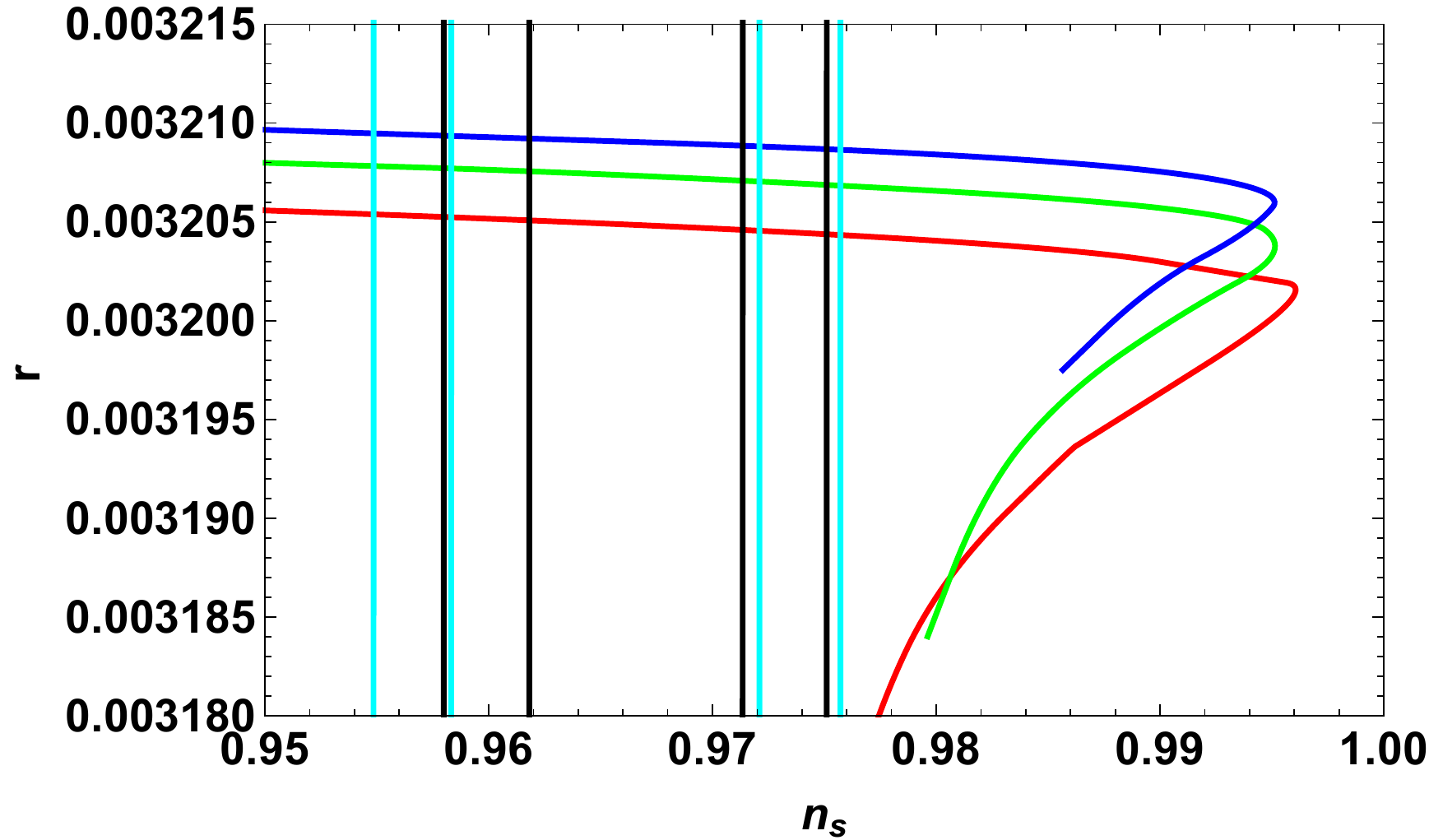}}
\caption{\small In case of cubic dissipation regime figure (a) and (b) represents plot between $n_s$ and $r$ for cases $n=4$ and $6$, respectively The magenta lines are for the Planck2018+lensing data, whereas the black solid lines correspond to Planck 2018+lensing+BK15+BAO \cite{aghanim2018planck} data. The pivot scale is fixed at usual $k=$ 0.002Mp$c^{-1}$. The green, red and blue curves depicts $N=50, 60$ and $70$, respectively.
}
\label{cubic4}
\end{figure}


In WI, the amplitude of scalar perturbations gets enhanced due to the additional contribution of thermal fluctuations in the inflaton field. As a result, the scalar power spectrum $P_\zeta$ takes the following form~\cite{dimop-owen}:
\begin{equation} \label{ps}
    P_\zeta = \left( \frac{H^2}{2\pi \dot\phi}\right)^2 \left(1 + W \right) \ , \quad \text{where} \quad W=\frac{T}{H}\frac{2\sqrt{3}\pi Q}{\sqrt{3+4\pi Q}} 
\end{equation}
%
As  mentioned earlier that we can express $Q_\ast$ in terms of other parameters, using that we can further find out the functional dependence of the power spectrum $P_\zeta$ on $V_0, n,\lambda,N, g_\ast$ and coupling constants ($C_\phi$ or $C_T$) which are completely arbitrary at this stage. By using Eqs. (\ref{slow-rolleqns}), (\ref{phit}) and (\ref{H}), we can express $P_\zeta$ at the pivot value as
\begin{equation}
P_{\zeta\ast} = \frac{H_\ast ^2 \,(Q_\ast+1)^2\, \phi_\ast^{2-2 n} \left(\frac{3 \times 2^{3/4} \times\,\sqrt[4]{5} \,\sqrt{\pi }\, \sqrt{\lambda }\, \sqrt{n} \, Q_\ast ^{5/4}\,
   \phi_\ast ^{\frac{n-1}{2}}}{\sqrt[4]{g_\ast}\, \sqrt{H_\ast}\, \sqrt{Q_\ast+1}\, \sqrt{4 \pi  Q_\ast+3}}\,+\,1\right)}{4\, \pi ^2\, \lambda ^2\,
   n^2} \ .
\end{equation}
The above expression, in general, is valid for both linear and cubic case. However, it depends on $C_\phi$ or $C_T$ in both cases through $Q_\ast$ and $\phi_\ast$. Here, let us emphasize that upon increasing $n$ and $\lambda$, $P_{\zeta\ast}$ also increases but remains inert to $V_0$.

Note that in the limit of $T\rightarrow 0$, the power spectrum takes it usual form of the CI. The quantity which represents the scale-dependence of the power spectrum dubbed spectral-index $n_s$, is defined as:
\begin{equation}
    n_s-1 \equiv \frac{d \ln P_{\zeta}}{H dt } \ .
\end{equation}
By taking the time derivative of Eq.(\ref{ps}), the spectral index $n_s$ can also be expressed as
\begin{equation}
n_s-1 =  \frac{1}{H^2}\left(\frac{6T\dot{H}+2H(2(1+W)\dot{H}+\dot{T})+H^2\dot{W}}{2T+H(1+W)}-\frac{2H\ddot{\phi}}{\dot{\phi}} \right)   \ .
\end{equation}
From Eq. (\ref{slow-rolleqns}), (\ref{slow-roll param}), (\ref{H}) and (\ref{T}), we obtain an explicit expression for $n_s$
in case of the linear dissipation regime:
\begin{align}\label{ns_linear}
    n_s = -\Big(3 \times 2^{3/4} \sqrt[4]{15} \sqrt{\pi } \sqrt{\lambda } \sqrt{n} \, Q^{5/4}
   (4 \pi  Q+3) \, \text{x1} \, \phi ^{n/2} \exp(\lambda  \phi ^n /4)+\text{x3}+\text{x4}\\ 
   -4 \sqrt{Q+1}\, (4 \pi  Q+3)^{3/2} \sqrt[4]{V_0} \, \text{x2}\, \sqrt[4]{g_\ast}\Big)\Big/ \Big(4 \phi ^2 (Q+1) (4 \pi  Q+3) \, \text{x5}\Big) \nonumber
\end{align}
where,
\begin{align}
x1= 3 \lambda ^2 n^2 \phi ^{2 n}+6 \lambda  (n-1) n \phi ^n-4 \phi ^2 (Q+1), 
\end{align}
\begin{equation}
x2= -\lambda ^2 n^2 \phi ^{2 n+\frac{1}{2}}-2 \lambda  (n-1) n \phi ^{n+\frac{1}{2}}+\phi ^{5/2} (Q+1), \\
\end{equation}
\begin{equation}
x3= \frac{4\times 3^{2/3} \pi ^2 \lambda \, n \,Q^{4/3} (Q+1)^{3/2} (4 \pi  Q+3)^{3/2} \sqrt[4]{V_0}\, g_\ast ^{5/4}\, \phi ^{n+\frac{31}{6}}\, \exp(\lambda  \phi
   ^n /3) \left(n \left(\lambda \, \phi ^n-6\right)+14\right)}{3^{2/3}\, \pi ^2\, \phi ^{14/3}\, \sqrt[3]{Q} \,(Q+1)\, (5 Q+2)\, g_\ast \,\exp(\lambda  \phi
   ^n /3)-180 \,\lambda ^2 \,\sqrt[3]{V_0} \,C_T^{4/3}\, \phi ^{2 n}}, \\
   \end{equation}
   \begin{align}
x4&=& \frac{9 \sqrt[4]{\frac{5}{2}} 3^{11/12} (\pi ^{5} \lambda ^{3} n^{3} Q^{9/6})^{1/2}(Q+1) (Q (4 \pi  (3 Q+1)+11)+5) g_\ast \phi ^{\frac{3
   n}{2}+\frac{14}{3}} e^{\frac{7 \lambda  \phi ^n}{12}}}{3^{2/3} \pi ^2 \phi ^{14/3} \sqrt[3]{Q}
   (Q+1) (5 Q+2) g_\ast \exp(\lambda  \phi ^n /3 )-180 \lambda ^2 \sqrt[3]{V_0} C_T^{4/3} \phi ^{2 n}} \times\left(n \left(\lambda  \phi ^n-6\right)+14\right),
   \end{align}
   \begin{equation}
x5= \sqrt[4]{g_\ast} \sqrt{\phi } \sqrt{Q+1} \sqrt{4 \pi  Q+3} \sqrt[4]{V_0}+3\ 2^{3/4} \sqrt[4]{15} \sqrt{\pi } \sqrt{\lambda } \sqrt{n}
   Q^{5/4} \phi ^{n/2} \exp(\lambda  \phi ^n/4).
\end{equation}
Again for the cubic case one gets:
\begin{align} \label{ns_cubic}
n_s =\Big(4 \sqrt[4]{g_\ast} \sqrt{Q+1} (4 \pi  Q+3)^{3/2} \sqrt[4]{V_0} \left(\phi ^{5/2} \text{x4}-\lambda ^2 n^2 \sqrt{V_0} \phi ^{2 n+\frac{1}{2}}\right)\\ \nonumber
   + 3/ 2^{3/4} \sqrt[4]{15} \sqrt{\pi } \sqrt{\lambda } \sqrt{n} \sqrt[4]{Q} \phi ^{n/2} \exp(\lambda  \phi ^n/4) \\ \nonumber
    \left(-3 \lambda ^2 n^2 Q (4 \pi  Q+3) \sqrt{V_0} \phi ^{2 n}-6 \lambda  (n-1) n Q (4 \pi  Q+3) \sqrt{V_0} \phi ^n+\phi ^2
   \text{x2}\,\, \right)-\text{x3}\Big)\\ 
   \Big/ \Big(4 \phi ^2 (Q+1) (4 \pi  Q+3) \sqrt{V_0} \,\,\text{x1}\Big) \nonumber
\end{align}
where,

\begin{eqnarray}
x1&=&\sqrt{\phi } \sqrt{Q+1} \sqrt{4 \pi  Q+3} \sqrt[4]{V_0} \sqrt[4]{g_s}+3\times 2^{3/4} \sqrt[4]{15} \sqrt{\pi } \sqrt{\lambda } \sqrt{n}\,  Q^{5/4} \phi
   ^{n/2} \exp(\lambda  \phi ^n/4), \\
x2&=& 3 \sqrt{3} (Q (12 \pi  Q+4 \pi +11)+5) Q'(t) \exp( \lambda  \phi ^n/2)+4 Q (Q+1) (4 \pi  Q+3) \sqrt{V_0}, \\
x3&=& 8 \sqrt[4]{g_s} \lambda  (n-1) n \sqrt{Q+1} (4 \pi  Q+3)^{3/2} V_0^{3/4} \phi ^{n+\frac{1}{2}}, \\
x4&=& 2 \sqrt{3} \exp(\lambda  \phi ^n/2) Q'(t)+(Q+1) \sqrt{V_0}.
\end{eqnarray} 
One can show that the above expressions (\ref{ns_linear}) and (\ref{ns_cubic}) recovers its usual form: $n_s-1 = -6\epsilon+2\eta$  in case of CI i.e.when $Q=0$.

The parametric dependence of on $n_s$ for the linear case is shown in fig. (\ref{linear3}) in which we depict two separate cases $n=4$ and $6$ for a range of parameters $\lambda$, $Q$ and $C_T$. In that figure, one can see that in both cases, the large values of $C_T$ corresponds to large values of $n_s$ and vice-versa. Similarly, for the cubic case shown in fig. (\ref{cubic3}), the similar behaviour is observed as in this case also large values of $C_\phi$ corresponds to large values of $n_s$ and vice-versa. 

From figures (\ref{linear3}) and (\ref{cubic3}) it is evident that $n=4$ case, which is strictly prohibited in the CI regime as shown in \cite{geng2017observational}, is completely feasible in the WI for both types of couplings: linear and cubic. Thus, in our following analysis we only consider $n=4$ and $6$ cases to find which one suits better and to check the compatibility of $n=4$ case with other observations. 


\subsection{Tensor perturbations}

Unlike the scalar power spectra $P_\zeta$, the dissipation of scalar field into radiation does not (directly) affect the tensor power spectra $P_T$ (it affects $P_T$ through $Q$ and $\phi$) dependence and therefore has the standard form,
\begin{equation} \label{P_T}
    P_{T\ast}= \frac{2}{\pi ^2}\left(\frac{H_\ast}{M_{pl}}\right)^2 \ .
\end{equation}
%
The tensor-to-scalar ratio $r$ is defined as the ratio of tensor to scalar spectra as 
\begin{equation} \label{r}
    r \equiv \frac{P_{T\ast}}{P_{\zeta \ast}} =  \frac{2}{\pi ^2}\left(\frac{H_\ast ^2}{2.1\times 10^{-9}}\right) \ .
\end{equation}
Note that here we have used the pivot-scale measurement, $P_{\zeta\ast} = 2.1\times 10^{-9}$.

\begin{figure}[t!] \label{ns-l figure}
\centering
\subfigure[]{%
\label{fig:first}%
\includegraphics[height=2.45in,width=2.65in]{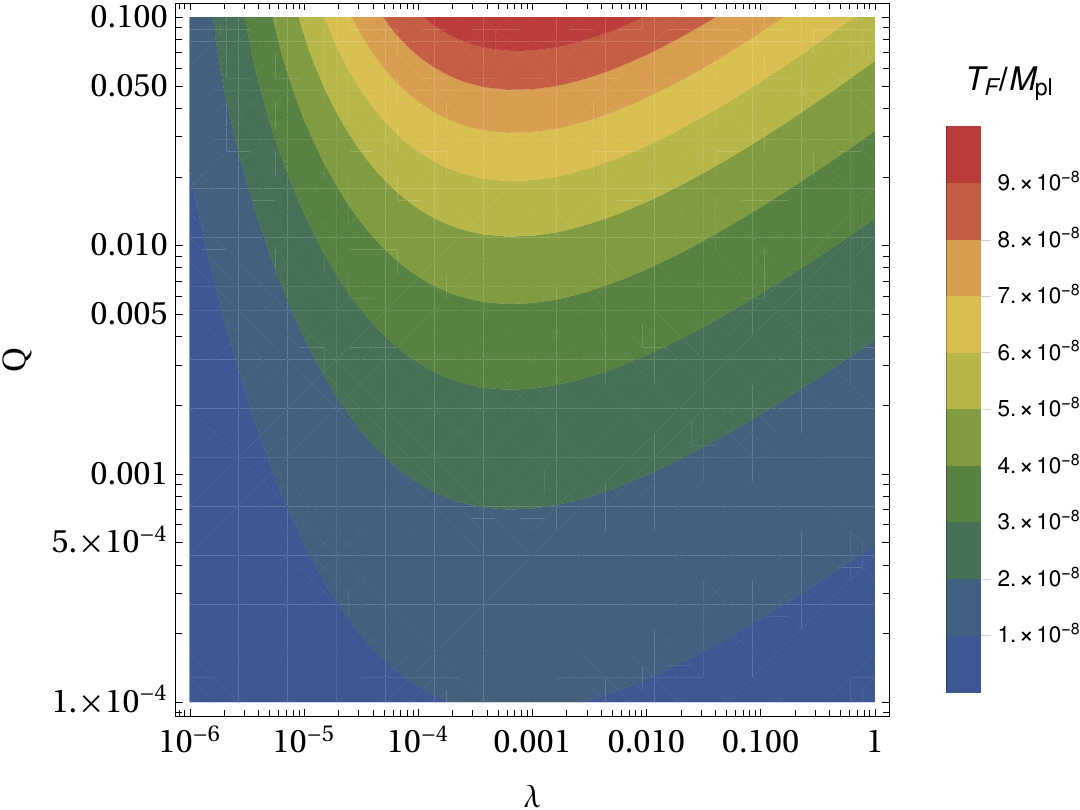}}%
\qquad \qquad 
\subfigure[]{%
\label{fig:second}%
\includegraphics[height=2.45in,width=2.65in]{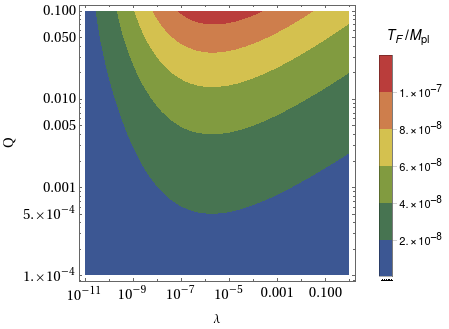}}%
 \caption{\small Plots (a) and (b) show the contours obtained between constant coupling parameter $Q$ and $\lambda$ for $n=4$ and 6, respectively. Each contour represents a value of freezing temperature $T_F $\cite{safiabaryog} as shown in the bar (on right side).}
    \label{bar}%
\end{figure}

At the pivot-scale, $H_\ast \simeq 10^{-6}M_{pl}$, that gives $r \simeq \mathcal{O}(10^{-3})$ which easily satisfies the Planck 2018 bound, $r<0.07$ \cite{aghanim2018planck}. Now, in order to check for the allowed parametric region between $n_s$ and $r$, we first vary $\lambda$ for both linear (Fig. (\ref{nssr})) and cubic ( Fig.\ref{cubic2}) cases. We find that $\lambda$ though has little dependence on $r$ (this is expected as $H$ is also weakly dependent on it) but can significantly affect the spectral index $n_s$. Alternatively, from the bounds on $n_s$ one can get the desired range of $\lambda$ which will naturally give rise to small $r$. Moreover, we also depict the parametric dependence of $n_s$ and $r$ at the pivot-scale for linear and cubic case. For linear case, we vary $C_T \in [10^{-5},10^{-1}]$ for $n=4$ and $6$ for different values of the e-foldings, $N=50,60$ and $70$ which corresponds to blue, red and green curves, respectively. The sudden increase in the slope of $n_s$ vs. $r$ after a certain limit is due to the fact that thermal fluctuations becomes dominant over quantum fluctuations. Similarly, for the cubic case, we vary $C_\phi \in [10^{6},10^{7}]$ for the same values of $n$ and $N$. Interestingly, we find that $n_s$ might take values greater than unity in case of $n=4$. And this results into  blue scalar spectrum which has its implications for the formation of primordial black holes. However for $n=6$, the spectral index, $n_s$ lies in the allowed red spectrum region, as usual. 
Thus, we find that the model corresponding to $n=4$, incompatible in case of CI, has rather interesting implications in the framework of WI. 
Let us make a remark on the possibility of baryogenesis in the paradigm of warm quintessential inflation under consideration.
As  stated earlier, in this case,  the field enters the kinetic regime after inflation ends and stays there for a long time and provides with an arena for producing baryon asymmetry  in the  thermal equilibrium a la spontaneous baryogenesis, $  \mathcal{L}_{\scriptsize \mbox{eff}} = {\lambda'}\partial_\mu \phi J_B^\mu$/M,
where $J_B^\mu$ is non-conserved baryonic current, $M$ is a cut scale and   $\lambda'$ is coupling parameter 
\cite{safiabaryog}. In this framework, we can analytically estimate the freeze out values of asymmetry, $\eta_F$ and the corresponding temperature\footnote{As a first check, we assume here coupling  to be constant.} as a function of $Q$, see Fig. (\ref{bar}), details are deferred to our future investigations \cite{gango}.

\section{Relic gravitational waves and observational constraints}
One of the important features of inflationary scenarios includes  the quantum mechanical production
of relic gravity waves whose experimental detection would provide a clean signal for falsification of the paradigm. The framework of quintessential inflation has a generic feature, namely, it is necessarily followed by kinetic regime which gives rise to blue tilt in the spectrum of
 gravity waves. The detection of stochastic background of these waves is a challenge to the forthcoming gravitational
wave missions. 
In this section, we study the production and post inflationary evolution of relic gravity waves in the set up of warm inflation.\footnote{For some recent developments of GW production, readers are urged to go through \cite{suk1, fig1, bernal}.}

 
The GW are tensor fluctuations $h_{ij}$ in the `linearized' space-time metric: $ds^2 = dt^2-a(t)^2(\delta_{ij}+h_{ij})dx^i dx^j$ such that  $h_{ij}$ satisfies the transverse-traceless conditions, i.e., $\partial_i h_{ij}=0=h^i_i$. 
The tensor fluctuations satisfy the Klein-Gordon equation in vacuum, $\Box h_{ij}=0$ and has the plane wave solution in the form,
\begin{equation}
    h_{ij}(t,{\bf{x}}) = \frac{1}{(2\pi)^{3/2}}\int d^3 k  \sum\limits_{\alpha = +,\times}\epsilon^\alpha_{ij}({\bf{k}})h^\alpha_{\bf{k}}(t,{\bf{k}})e^{{i\bf{k}\cdot\bf{x}}} \ ,
\end{equation} 
where $\epsilon^\alpha_{ij}$ is the polarization tensor symmetric in $i,j$ and $\alpha$ represents two polarization states: $+,\times$. By redefining the time-coordinate to conformal time $d\tau=\frac{dt}{a(t)}$, the GW equation in the momentum space can be written as
\begin{equation}
    {h^\alpha_{\bf k}}''(\tau)+2 \mathcal{H} {h^\alpha_{\bf k}}'(\tau)+k^2h^\alpha_{\bf k}(\tau) = 0 \ ,
\end{equation}
where $k\equiv |\bf{k}|$, $'\equiv \frac{d}{d \tau}$ and $\mathcal{H} = \frac{a'}{a}$ is the conformal Hubble parameter. The spectrum of primordial GW is defined as an interval in its energy density per logarithmic frequency per critical density $\rho_c(\tau) \equiv \frac{3H^2(\tau)}{8\pi G}$,
 \begin{equation} \label{omegagw}
     \Omega_{\mbox{\scriptsize GW}}(\tau,k) \equiv \frac{1}{\rho_c(\tau)}\frac{d \rho_{\mbox{\scriptsize GW}}}{d \ln k}  \ , 
 \end{equation}
where $\rho_{\mbox{\scriptsize GW}}$ is the GW energy density, and is expressed as
 \begin{eqnarray} \label{rgw}
     \rho_{\mbox{\scriptsize GW}} &=& \frac{1}{16 \pi G} \int_0^k \frac{d^3 k}{(2\pi)^3}\frac{k^2}{a^2} \sum\limits_{\alpha =+,\times} |h^\alpha_k|^2 \nonumber \\ &=&  \frac{k^5}{160 \, \pi^3 G a^2}  \sum\limits_{\alpha =+,\times} |h^\alpha_k|^2 \ .
 \end{eqnarray} 
By using Eqs. (\ref{omegagw}) and (\ref{rgw}), the GW energy density parameter at present epoch $(t=t_0)$ can be expressed in terms of tensor power spectrum $\Delta^2_T(k)$ as
\begin{equation}
     \Omega_{{\mbox{\scriptsize GW}},0} = \frac{1}{12}\left( \frac{k^2}{a_0^2 H_0^2}\right)\Delta^2_T(k) \ , \quad \mbox{where} \quad \Delta^2_T(k) \equiv \frac{k^3}{\pi^2}\sum\limits_{\alpha =+,\times} |h^\alpha_k|^2 \ .
 \end{equation}
In order to estimate the GW spectrum $\Delta^2_T(k)$ at present epoch from the primordial spectrum, let us use the relation 
 \begin{equation}
     \Delta^2_T(k) = P_T(k)T^2(k) \ ,
 \end{equation}
where $P_T(k)$ is the primordial tensor power spectra and is already defined in Eq. (\ref{P_T}) whereas $T^2(k)$ is the transfer function which determines the change or evolution in the shape of $P_T(k)$ as the Universe goes through different eras which leads to a change in the profile of the observed power spectra \footnote{Note that the factor $1/2$ corresponds to the root-mean-square value of $\Omega_{{\mbox{\scriptsize GW}},0}$. }
\begin{equation} \label{transfer}
    T^2(k) = \frac{1}{2} \left( \frac{a(k)}{a_0}\right)^2 \ ,
\end{equation}
where $a(k)$ represents the scale-factor when a particular $k$ mode crosses the horizon. The rate of expansion of the Universe during different eras is given by the Hubble parameter as \cite{safiabaryog}
 \begin{equation} \label{hubble parameter}
     H(a) = H_0 \sqrt{\Omega_\phi(a)+\Omega_{r,0}\left(\frac{g_\ast}{g_{\ast 0}}\right)\left(\frac{g_{\ast s}}{g_{\ast s 0}}\right)^{-4/3}\left(\frac{a}{a_0}\right)^{-4}+\Omega_{m,0}\left(\frac{a}{a_0} \right)^{-3}} \ ,
 \end{equation}
where $g_{\ast s}$ denotes the effective relativistic degrees of freedom for entropy density. From the above Eq. (\ref{hubble parameter}), one can determine the scale factor at the moment when different $k-$ modes crosses the horizon in different eras

\begin{equation} \label{a(k)}
  a(k_\ast) =
    \begin{cases}
    a_r \left( \frac{k_r}{k}\right)^{1/2} , & \text{Kinetic regime} \\
     \left(\frac{g_\ast}{g_{\ast 0}}\right)^{1/2}\left(\frac{g_{\ast s}}{g_{\ast s 0}}\right)^{-2/3}\frac{H_0 \Omega_{r,0}^{1/2}}{k}, & \text{RDE} \\
       \Omega_{m,0}\frac{H_0^2}{k^2}, & \text{MDE} \\
    \end{cases}       
\end{equation}
Here, we have used $a_0 =1$ and relation $k_\ast=a_\ast H_\ast$. Here, $a_r$ represents the scale factor at the beginning of the radiation era and $k_r$ represents the mode which enters inside the horizon at that epoch. 
One can also find out the amount of energy GW carries at present epoch which had entered the horizon in different epochs. From Eqs. (\ref{transfer}), (\ref{hubble parameter}) and (\ref{a(k)}) we get,
\begin{eqnarray}
 \Omega_{{\mbox{\scriptsize GW}},0}^{KD} &=& \Omega_{{\mbox{\scriptsize GW}},0}^{RD} \left( \frac{k}{k_r}\right) \ , \quad k \in (k_r,k_{end}] \label{omega-gwkd}\\
\Omega_{{\mbox{\scriptsize GW}},0}^{RD} &=& \frac{1}{6\pi^2}\frac{\Omega_{r,0}H_{\ast}^2}{M_{pl}^2}\left(\frac{g_\ast}{g_{\ast 0}}\right)\left(\frac{g_{\ast s}}{g_{\ast s 0}}\right)^{-4/3} \ , \quad k \in (k_{eq},k_r] \label{omega-gwrd} \\
\Omega_{{\mbox{\scriptsize GW}},0}^{MD} &=& \frac{1}{6\pi^2}\frac{\Omega_{m,0}^2H_{\ast}^2}{M_{pl}^2}\frac{H_0^2}{k^2} \ , \quad  k \in (k_0,k_{eq}] \ . 
\label{omega-gw} \end{eqnarray}
where $k_0,k_r,k_{eq}$ and $k_{end}$ are the modes which enters inside the horizon at present epoch, radiation dominated era, equality epoch and at the end of inflation, respectively. Note that the modes which enters during the RDE the GW amplitude of them is independent of their frequency.

\begin{figure}
    \centering
   \includegraphics[height=2.5in,width=3in]{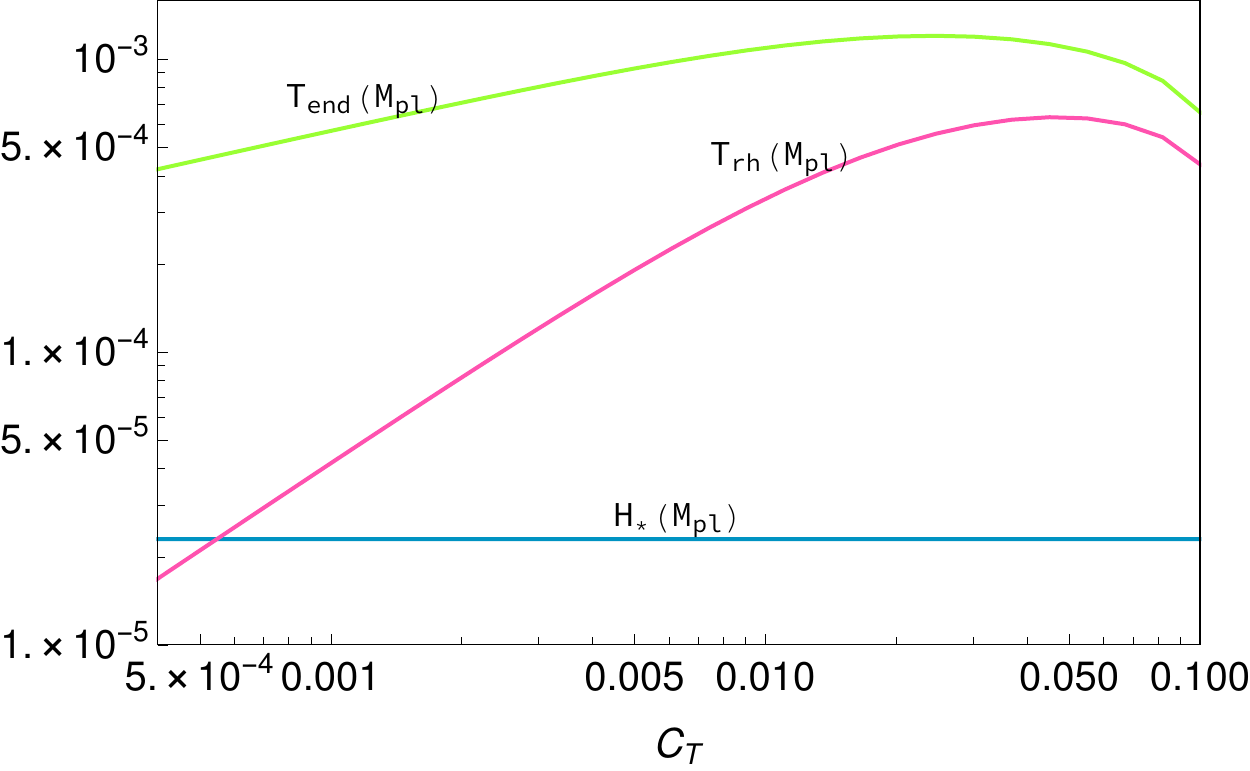}
    \caption{\small Variation of $H_\ast$, $T_{rh}$ and $T_{end}$ with $C_T \in [5\times 10^{-4},0.1]$ for $n=6$ and $\lambda=10^{-4}$.}
    \label{H_ast}
\end{figure}

Let us now briefly describe the evolution of radiation and field densities in post-inflationary era and their impact on the production of relic GWs. For this, we make an assumption that after inflation, the radiation and field now evolve independently.
\begin{equation} \label{rhor}
\rho_r = \left(\frac{a_{\text{end}}}{a}\right)^4 \rho _{r(\text{end})} \ , \quad \text{where} \quad \rho_{r({\text{end}})} = \frac{Q_{\text{end}} V_{\text{end}}}{2 \left(Q_{\text{end}}+1\right)} \ ,
\end{equation}
whereas, 
\begin{equation} \label{rhophi}
\rho _{\phi } = \left(\frac{a_{\text{end}}}{a}\right)^6 \rho _{\phi(\text{end})} \ ,  \quad \text{where} \quad \rho _{\phi(\text{end})} = \frac{\left(3 Q_{\text{end}}+4\right) V_{\text{end}}}{3 \left(Q_{\text{end}}+1\right)} \ .
\end{equation}
Soon after inflation ends, scalar field enters the kinetic regime; field energy density is much smaller than $\rho_r$ which however redshifts slower than $\rho(\phi)\sim a^{-6}$ and at a particular epoch takes it over.
The epoch when both energy densities becomes equal i.e. $\rho_r =\rho _{\phi } $ would be  referred to as {\it reheating}. The scale factor at reheating can be estimated by equating Eq. (\ref{rhor}) and (\ref{rhophi}) as,
\begin{equation} \label{arh}
a_{rh} = a_{\text{end}} \left(\frac{\rho _{\phi (\text{end}) }}{\rho _{r(\text{end})}}\right)^{1/2} \ .
\end{equation}
Also, the reheating temperature $T_{rh}$ can be expressed as 
\begin{equation} \label{Trh}
T_{rh} = T_{\text{end}} \left(\frac{a_{\text{end}}}{a_{rh}}\right) \ .
\end{equation}
Let us emphasize that $T_{end}$ for WI is usually larger than its counter part in framework of CI, for e.g., by taking $ \{n,\lambda,V_0,C_T(C_\phi) \}=\{ 6,10^{-3},10^{-10},10^{-4.5}(10^{4.5})\}$, $T_{end} \simeq 10^{-4}M_{pl}$, which is around one order of magnitude larger than that  in the paradigm of CI. Also, for the same value of model parameters we find $a_{rh}=96.1\, a_{end}$ and $28.5 \, a_{end}$ and $T_{rh} =1.24\times 10^{-6}M_{pl}$ and $ 7.71 \times 10^{-6} M_{pl}$ for the linear and cubic case, respectively. Hence we find that the reheating temperature $T_{rh}$ in WI is  larger than that of the CI value (see fig. (\ref{H_ast}).
The transition frequencies $f=k/2\pi$ can be determined from the relation $k =aH$ as
\begin{eqnarray} 
\text{at present:} \quad f_0 = \frac{a_0 H_0}{2\pi} \simeq 3.5 \times 10^{-19} \text{Hz} \ , \nonumber \\
\text{at matter-radiation equality:} \quad f_{eq} = \frac{a_{eq} H_{eq}}{2\pi} \simeq 1.6 \times 10^{17} \text{Hz} \ , \nonumber \\
\text{at radiation-dominated era:} \quad f_r = \frac{a_{r} H_{r}}{2\pi} \simeq 3.896 \times 10^{11} T_{rh} \, \text{Hz} \ , \nonumber \\
\text{at the end of inflation :} \quad f_{end} = \frac{H_{end}}{2\pi}\left(\frac{T_0}{T_{end}}\right) \text{Hz} \ ,
\label{transitionfrequencies}
\end{eqnarray}
where $T_0 = 2.73K$ and we have used $g_\ast =106.75$.
By using Eqs. (\ref{a(k)}-\ref{transitionfrequencies}), GW energy densities can be expressed as
\begin{equation}
\Omega_{{\mbox{\scriptsize GW}},0}^{RD}h^2 =      \frac{(2.08\times 10^{-3})\, h^2\,  V(\phi_\ast) \, \Omega _{r,0}}{M_{\text{pl}}^4} \ ,
\end{equation}
\begin{equation}
 \Omega_{{\mbox{\scriptsize GW}},0}^{KD}h^2 =    \frac{(1.57\times 10^{-14}) \, h^2\, \sqrt[4]{Q_{end}+1} \, V(\phi_\ast) \,\Omega
   _{r,0} f}{\sqrt[4]{Q_{end}V_{end}} M_{\text{pl}}^3} \left(\frac{a_{\text{rh}}}{a_{\text{end}}} \right) \ ,
\end{equation}
\begin{equation}
  \Omega_{{\mbox{\scriptsize GW}},0}^{MD}h^2  = \frac{h^2 f_0^2 V(\phi_\ast) \Omega _{\text{m,0}}^2}{18 \pi ^2 f^2 M_{\text{pl}}^4} \ .
\end{equation}
It has been pointed out in \cite{sami-sahni} that it is the transition from inflationary era to kinetic regime which gives the largest contribution to the energy of relic GW. Therefore, the amplitude of GW from the transition can be expressed as
\begin{equation}
     \Omega_{{\mbox{\footnotesize GW}},0}^{max} = \Omega_{{\mbox{\scriptsize GW}},0}^{RD}\left(\frac{a_r}{a_{end}}\right) \ .
\end{equation}

\begin{figure}
 \centering
    \includegraphics[height=2.5in,width=3in]{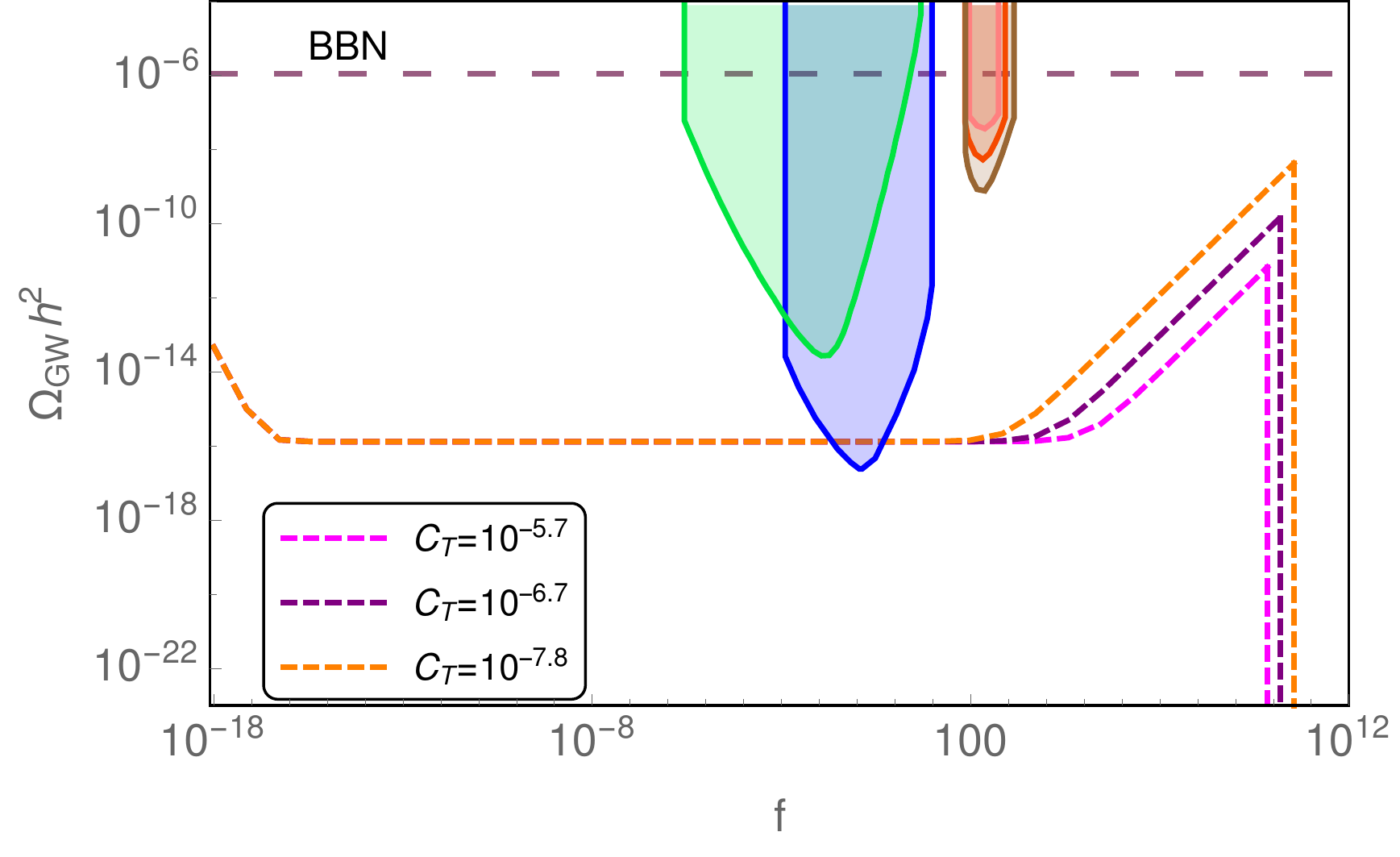}
    \includegraphics[height=2.5in,width=3in]{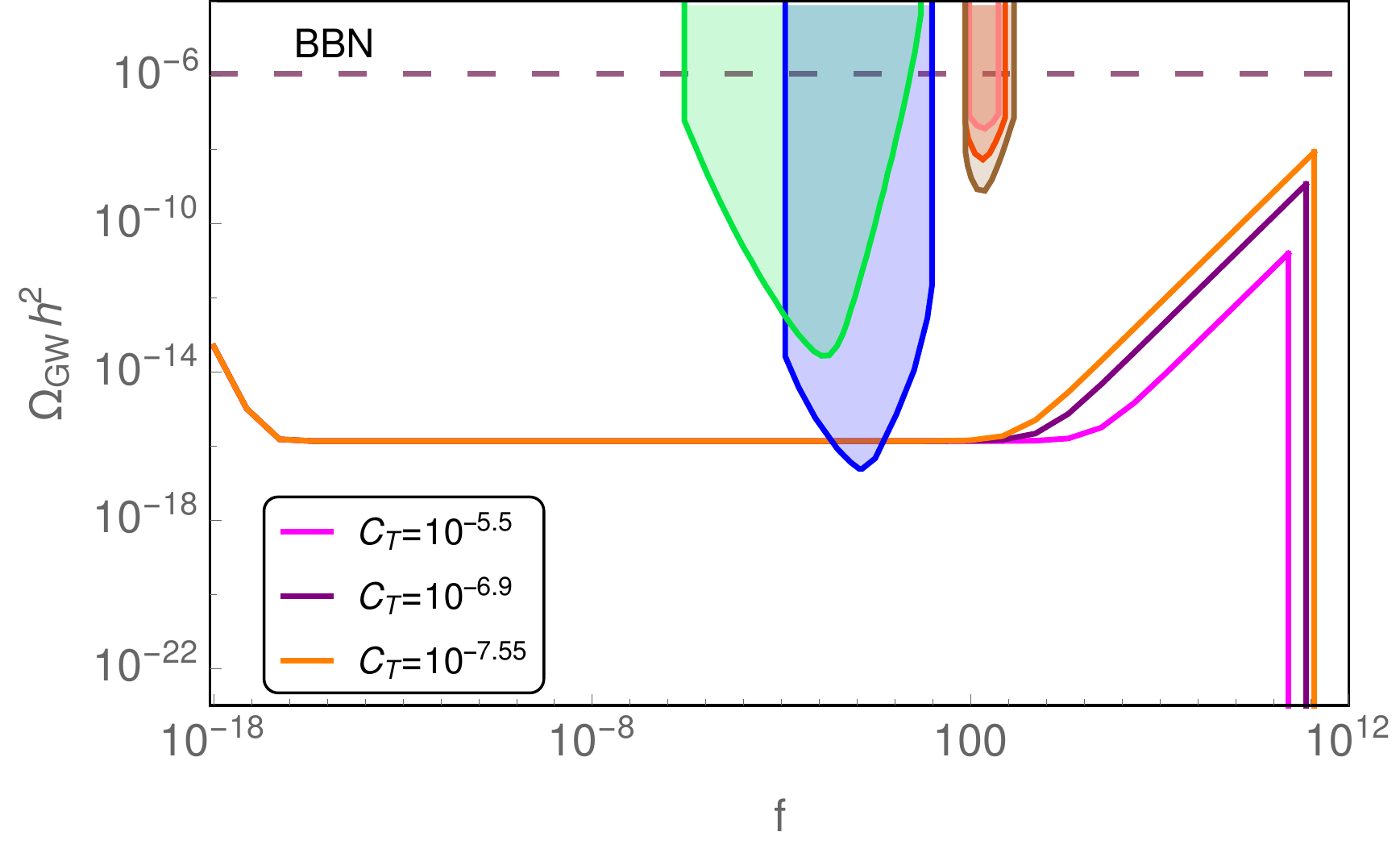}
    \caption{\small{The relic gravitational wave spectrum for linear dissipation; left figure corresponds to the $n=4$ and right figure $n=6$ cases respectively. Also the green, blue and red patches corresponds to the LISA, DECIGO and LIGO (different stages) bounds.}}
    \label{gw4}
\end{figure}

The big bang nucleosynthesis (BBN) puts strong constraint on this peak value of relic GW: $\Omega_{{\mbox{\scriptsize GW}},0}h^2< 1.12 \times 10^{-6}$. 
 Let us note that
 the gravitational particle production though is a Universal process  but challenged by BBN constraint which is related to the inefficiency of the process or relative smaller value of radiation energy density produced in the this process.
 
 The network of ground based detectors such as KAGRA \cite{kagra}, advanced LIGO \cite{abbott} and Virgo \cite{virgo} would be able to probe GW background up to $\Omega \sim 10^{-9}$ for frequencies around $10^2$ Hz  as their proposed sensitivities are reached in near future.
 As for the proposed satellite detectors LISA \cite{lisa} and DECIGO \cite{kawamura}, they would probe lower frequency band $10^{-1}- 10^{-3}$ Hz, whereas SKA \cite{ska} will access GW background with frequencies of the order of $10^{-8}$ Hz. 
 
 We have plotted our numerical results
 in Fig. (\ref{gw4}) for $n=4$ and $n=6$ which shows the spectra of stochastic background of relic gravity waves for different values of $C_T$. Let us note that, $\Omega^{RD}_{GW,0}$ $\&$ $\Omega^{MD}_{GW,0}$ depend upon $H_\ast$ (see, Eqs. (\ref{omegagw}) and (\ref{rhor})) which is weakly dependent on $C_T$ (see Fig. (\ref{H_ast})) thereby this part of the spectrum is practically same as in case of CI. 
 However, $\Omega^{KD}_{GW,0}$ is inversely proportional to frequency at the commencement of radiative regime, $f=f_r$ (see Eq. (\ref{omega-gwkd})) which, in turn, is directly proportional to the reheating temperature $T_{rh}$(see, Eq. (\ref{transitionfrequencies})). As for $ T_{rh}$, it increases with the increase in $C_T$ (see Fig.(\ref{H_ast}))
 \footnote{$T_{end}$ also has similar behavior. }; as a result, the GW spectrum gets modified:(1) In the WI scenario, the blue spectrum gets shifted towards higher frequency region and (2).
 The GW amplitude is slightly suppressed compared to the case of cold inflation\footnote{For smaller values of $C_T$, our results move closer to the ones predicted by the framework of CI as it should be.}, (see, Fig. (\ref{gw4})). Let us emphasize that in the standard framework
 \footnote{It refers to usually considered model of inflation in which the inflaton enters the oscillatory regime after inflation and decays giving rise to (p)reheating.}, 
 inflation is followed by radiative regime and thus the blue  spectrum, $k\in (k_r,k_{end})$ (in the high frequency regime) is absent in that case. 
 In the scenario under consideration, the part of the spectrum corresponding to $k\leq k_r$, does not significantly differ from the one predicted by the standard inflationary scenario and might be probed in near future by the proposed GW missions
 \footnote{The DECIGO proposed sensitivity has little overlap with the predictions ($\Omega_{GW} \sim 10^{-16}$) for $k \leq k_r$ or $f\leq f_r$.}. 
 As for the blue spectrum $-$ the distinguished feature of the quintessential inflation, it appears in the high frequency band. The present  proposed sensitivities would miss this novel feature until they are improved to explore the high frequency region of the spectrum \cite{SRI,Cruise:2006zt,Cruise:2012zz,Arvanitaki:2012cn,Sabin:2014bua,Goryachev:2014yra,Chou:2016hbb,Robbins:2018thb,Ito:2019wcb,Dimow,bari} and this certainly throws a challenge to future missions on the hunt of relic gravity waves.  
 
  




\section{Conclusion}
\label{sec:con}
In this paper, we have investigated a class of inflationary models dubbed quintessential inflation, in presence of non-zero temperature which arises due to the field-radiation coupling during inflation. During inflation, the field dissipates into the radiation and causes the radiation to finally become dominant in the post inflationary era. Although, our study is based on the weak dissipative regime ($Q_\ast \leq 1)$ but we have shown that even if the coupling is small it still has  significant effects which not only allows us to make different theoretical predictions compared to that of the cold inflation scenario but also turns out to be more promising when it comes to satisfy the observational constraints. We have considered  the generalized exponential potential which gives rise to a successful inflationary scenario followed by the kinetic regime  before the the commencement radiative regime. Our investigations show that the field-radiation coupling allows us to satisfy the observational constraints for  values of parameters  ruled out in the cold inflationary scenario by Planck results. As briefly discussed, the coupling also, by maintaining thermal equilibrium during inflation, provides a natural setup for spontaneous baryogenesis to take place during kinetic regime. The freeze-out temperature, in this 
case, turns out to be proportional to the coupling parameter and always satisfies the constraint $T_F < T_{rh}$, detailed results would be reported elsewhere \cite{gango}.

Furthermore, we study the impact of the coupling on the production of the GW energy densities, the major contribution of which comes from the transition from inflation to kinetic regime. We find that  for the values of the model parameters which might be challenged by the BBN constraint, the presence of coupling    , gives rise to some relaxation which is in complete agreement with the bounds obtained from the Planck results on $n_s$ and $r$. Our numerical results are displayed in Fig. (\ref{gw4}) which shows that the blue spectrum, caused by the kinetic regime after inflation ends, gets shifted towards
higher frequencies and GW amplitude is slightly suppressed compared to the case of cold inflation. This behavior is attributed to the fact that the frequency of GW at the commencement of kinetic regime denoted by $k_r$ depends upon reheating temperature which is higher for larger values of $C_T$, see Fig. (\ref{H_ast}). As for the lower frequency part of the spectrum, $k\leq k_r$, it depends upon $H_\ast$ which is insensitive to $C_T$. This part of the spectrum repeats the predictions of quintessential inflation
in the cold background and might be probed by DECIGO in near future, see Fig. (\ref{gw4}). 
The blue spectrum in the high frequency region  is the distinguished feature of the quintessential inflation. The   proposed missions on the hunt of relic gravity waves are likely to miss the said novel feature until their sensitivities are improved to explore the high frequency region of the spectrum [\cite{SRI}-\cite{Ito:2019wcb}].

Another issue which might need a serious theoretical explanation is the observed anomaly at the low multipole in the CMB power spectrum as observed by Planck as well as WMAP. Many explanations \cite{lowl1}-\cite{lowl5} have been put forward and on that note it would be exciting to check if a warm quintessential inflation due to the presence of interactions, could explain such an effect in the large scales of the CMB.  Also, an interesting issue is associated with  inflaton as a composite scalar\cite{Bhattacharya:2018xlw}. 
Secondly, in cubic case when $\Upsilon$ is proportional to $T^3$, we found the blue scalar power spectrum i.e. $n_s \geq 1$ for some values of model parameters (see fig. (\ref{cubic4}))\footnote{Not to be confused with the blue spectrum due to post inflationary kinetic regime in the scenario of quintessential inflation.} which could be an interesting feature to study the formation of the primordial black holes. Last but not least, since, the model under consideration unifies early and late-time phases of cosmic acceleration, it also becomes important to study the effects of coupling on the late time evolution \cite{cope}. And we defer the study of these issues  to our future investigations.
\vspace{1cm}

\textbf{Acknowledgements:}
Work of MRG is supported by Department of Science and Technology, Government of India under the Grant Agreement number IF18-PH-228 (INSPIRE Faculty Award). 
MS is supported by
 the Ministry of Education and Science of the Republic of Kazakhstan, Grant No. 0118RK00935. SM is supported
 by the Ministry of Education and Science of the Republic of Kazakhstan, Grant No. BR05236322 and Grant N0 AP08052197. The work of MKS is supported by the Council of Scientific and Industrial Research (CSIR), Government of India. 


\end{document}